\documentclass[twocolumn]{aa}
\usepackage{graphicx}
\usepackage{txfonts}
\usepackage{comment}
\usepackage{float}
\usepackage{subfig}
\usepackage[dvipsnames]{xcolor}
\usepackage[utf8]{inputenc}
\usepackage[T1]{fontenc}
\usepackage{amsfonts}
\usepackage{listings}
\usepackage[colorlinks=true,allcolors=blue]{hyperref}
\usepackage{mathrsfs}
\usepackage{multirow}
\usepackage{tablefootnote}
\usepackage{subcaption}

\begin{document} 
\title{Grain alignment and dust evolution physics with polarisation (GRADE-POL)}
\subtitle{II. On the physical basis of Serkowski and super-Serkowski polarisation spectra}
\author{Le Ngoc Tram\inst{1}\thanks{Corresponding author: Le Ngoc Tram \newline \email{nle@strw.leidenuniv.nl}},
Thiem Hoang\inst{2,3},
B-G Andersson\inst{4},
Alex Lazarian\inst{5},
Lapo Fanciullo\inst{6},
Bao Truong\inst{3},
Archana Soam\inst{7}
}
\institute{
$^{1}$ Leiden Observatory, Leiden University, PO Box 9513, 2300 RA Leiden, The Netherlands \\
$^{2}$ Korea Astronomy and Space Science Institute, Daejeon 34055, Republic of Korea \\
$^{3}$ Department of Astronomy and Space Science, University of Science and Technology, 217 Gajeong-ro, Yuseong-gu, Daejeon, 34113, Republic of Korea \\
$^{4}$ McDonald Observatory, University of Texas at Austin, 2515 Speedway Boulevard, Austin, TX 78712, USA \\
$^{5}$ Department of Astronomy, University of Wisconsin-Madison, Madison, WI, 53706, USA \\
$^{6}$ National Chung Hsing University, 145 Xingda Rd., South Dist., Taichung City 402, Taiwan \\
$^{7}$ Indian Institute of Astrophysics, II Block, Koramangala, Bengaluru 560034, India
}
%
\titlerunning{Serkowski and super-Serkowski curves}
\authorrunning{Tram et al., 2026}
\abstract
{Optical-to-near-infrared interstellar polarisation, induced by aligned dust grains, generally follows a convex wavelength dependence, known as the Serkowski relation.  However, observations in the ultraviolet (UV) and at [mid-]infrared wavelengths have indicated that some of the spectra do not follow this relation. Specifically, about 25\% show an excess in the degree of polarisation at mid-UV wavelengths ($\lambda^{-1} > 3\,\rm \mu m^{-1}$), referred to as the super-Serkowski polarisation. For this study, we re-examined both the Serkowski and super-Serkowski spectra based on the joint effect of paramagnetic relaxation, known as the Davis-Greenstein (DG) and radiative torque (RAT) alignment.  We used the observational data for HD 30614, HD 204827, HD 37903 and HD 161056 to constrain our modelling. We examined two types of radiation fields: one derived from the scaled interstellar radiation field and the other originating from a B-type star. For the super-Serkowski spectra of HD 30614 and HD 204827, our model demonstrates that RAT alignment enhanced by radiation produced from a B-type star below the Lyman limit ($\lambda=912\AA$) can reasonably explain the observations and that a combination with the DG alignments results in a better fit for $\lambda^{-1}\geq 5.5\,\rm \mu m^{-1}$. For the Serkowski spectra in HD 37903 and HD 161056, only the RAT alignment by itself under the typical interstellar radiation field above the Lyman limit, within a typical cold neutral medium, can account for the observed spectra, with a combination of a very inefficient DG alignment. The capacity of our model to predict the starlight polarisation spectrum from infrared to far-UV is thus a promising tool for interpreting future missions that observe spectrophotometry in the UV bands.}

\maketitle
\section{Introduction}\label{sec:intro}
The polarisation of distant starlight was first discovered by \cite{1949Sci...109..166H} and \cite{1949Sci...109..165H}. This discovery revealed that interstellar dust grains are non-spherical and systematically aligned with an axis in the interstellar medium, presumably with magnetic fields. Soon after the discovery, many mechanisms were proposed to explain grain alignment, including paramagnetic relaxation (\citealt{DG1951}, hereafter the DG mechanism), gas flow \citep{Gold1952}, suprathermal rotation \citep{1979ApJ...231..404P}, and radiative torque (RAT) alignment \citep{1976Ap&SS..43..291D,1996ApJ...470..551D,lazarian2007}. Although the modern understanding favours the RAT mechanism for explaining the alignment of interstellar grains (see \citealt{2015ARA&A..53..501A,2015psps.book...81L} for reviews), the DG mechanism is expected to be important for small grains \citep{HoangLazMartin.2014}. The wavelength-dependent polarisation of starlight from UV to near-infrared (NIR) is crucial for constraining alignment mechanisms and dust properties (size distribution and composition).

Based on observations by G. V. Coyne and T. Gehrels \citep[e.g.][]{coyne1967}, K. Serkowski \citep{serkowski1971,1973IAUS...52..145S} formulated the homonymous empirical relation for its wavelength dependence (i.e. the polarisation spectrum) as
\begin{equation}
    p(\lambda)=p_{\rm max} \cdot \exp[-K \cdot \ln^2(\lambda_{\rm max}/\lambda)]{,}
\end{equation}
where $p_{\rm max}$ is the peak polarisation, occurring at the wavelength $\lambda_{\rm max}$, and $K$ (originally set to 1.15) controls the width of the curve. At that time, this formulation fit all known observations \citep{serkowski1971,1973IAUS...52..145S}.  Subsequent observations \citep{codina1976,wilking1980,wilking1982,1992ApJ...386..562W} showed deviations from the original Serkowski relation and, in particular, required a variable $K$-parameter, referred to as the Serkowski$-$Wilking relation \citep{1992ApJ...386..562W}:
\begin{equation} \label{eq:SW_law}
    K_{\rm VIR} = 0.01 \pm 0.05 + (1.66 \pm 0.09)\lambda_{\rm max}{.} 
\end{equation}
Ultraviolet (UV) range polarimetry taken with the Wisconsin Ultraviolet Photo-polarimetry Experiment (WUPPE) on board the Astro-1 and Astro-2 space shuttle missions and with the Faint Object Spectrograph (FOS) on the Hubble Space Telescope (HST) suggested a further modification for the UV \citep{1999ApJ...510..905M},
\begin{equation} \label{eq:modS_law}
    K_{\rm UV} = -0.59 \pm 0.21 + (2.56 \pm 0.38)\lambda_{\rm max}{,}
\end{equation}
showing a slightly steeper slope than the previous curve.

In addition, the WUPPE and HST/FOS observations, converting from near-IR to mid-UV, showed that the UV polarisation of the majority of the observed sight lines (19 out of 28) closely followed a short-wavelength extrapolation of the optical Serkowski curve. However, for 9 of the 28 lines of sight, a systematic polarisation excess from the Serkowski curve extrapolation was seen in the space-UV \citep[$\lambda^{-1}>3\,\rm \mu m^{-1}$][]{1995ApJ...445..947C,1999ApJ...510..905M}, referred to as super-Serkowski polarisation.

It is important to note that the Serkowski relation and its modifications are strictly empirical, and while theoretical studies have been able to reproduce many of its featured based on varying the alignment function for the underlying \citep[Mathis, Rumple \& Nordsieck; MRN][]{1977ApJ...217..425M} grain size distribution \citep[e.g.][]{kim1995,clayton2003}, it is only with the advent of modern grain alignment theory \citep{2007MNRAS.378..910L,2015ARA&A..53..501A} that a quantitative and physical understanding of the polarisation curve and its origins have been possible. Here, we focus on the relationship between the alignment of very small grains and the super-Serkowski UV polarisation, whose detailed physical origins are not yet fully understood.

In modern grain alignment theory \citep{2007MNRAS.378..910L,2015ARA&A..53..501A} grain alignment requires, to first order, that a paramagnetic grain is illuminated by radiation with wavelengths smaller than the grain diameter ($\lambda < 2a$).  The alignment can be enhanced by, for example, H$_2$ formation \citep{bga2013,hoang2015} and is balanced by gas-grain collisional randomisation and IR emission \citep{draine_1998}. Therefore, the alignment, and hence the polarisation spectrum, will depend on the grain size distribution and mineralogy (magnetic susceptibility), and on the physical conditions such as the spectral energy distribution (SED) of the illuminating radiation (including reddening), the gas density and temperature, and the magnetic field (a higher magnetic field strength could be responsible for the effective alignment of small grains; see \citealt{HoangLazMartin.2014}).

For neutral gas, numerical modelling of starlight polarisation in \cite{Hoang.2017SNe} demonstrated that the decrease in the alignment size of grains by stronger RATs due to stronger radiation fields can enhance the UV polarisation (see their Figure 6), which might reproduce the super-Serkowski spectra (see also \citealt{2022Ap&SS.367..127A}). However, due to the RAT condition for efficient grain alignment $\lambda < 2a$ for the typical intensity of the interstellar radiation field (ISRF), and because the ISRF spectrum has a cut-off point at the Lyman limit ($\lambda$ = 921\,\AA = 0.0912\,$\mu$m), this would require that extreme ultraviolet (EUV; $\lambda <$ 921\,\AA) radiation can propagate in the medium, and therefore that the medium is fully ionised.

\begin{figure}
    \centering
    \includegraphics[width=1.0\linewidth]{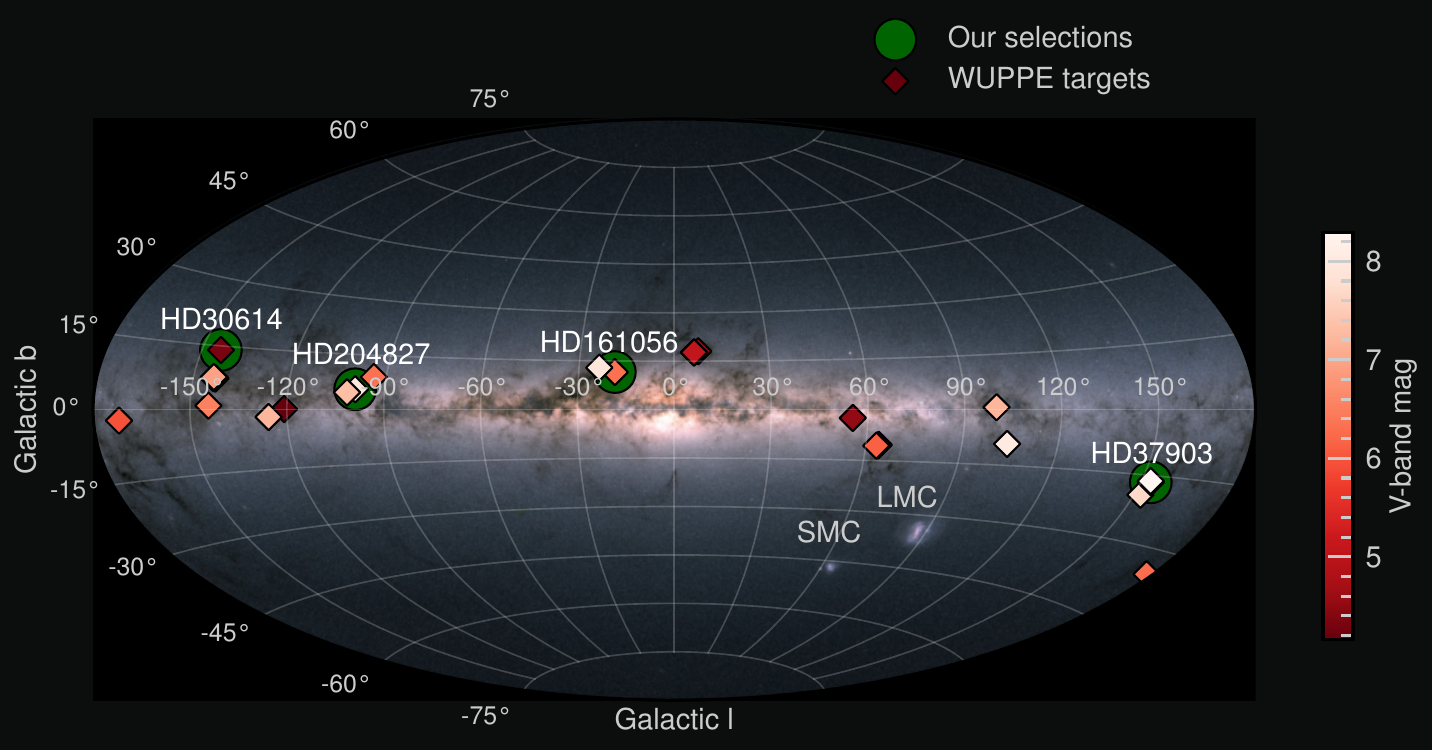}
    \caption{Locations of our selections of super-Serkowski polarisation spectrum (HD 37903 and HD 204827) and Serkowski polarisation spectrum (HD 30614 and HD 161056) within the WUPPE targets. The colour of the symbols indicates the magnitude in the V band towards these background stars. This figure was created with permission from \cite{10.1117/12.3065702}.}
    \label{fig:Wuppe_targets}
\end{figure}

The absolute alignment efficiency also depends on the magnetic susceptibility of dust grains \citep{HoangLaz.2016a,HoangLaz.2016b}. \cite{2018MNRAS.479.1685P} showed that a mixture composition (forsterite, Mg$_{2}$SiO$_{4}$, and enstatite, MgSiO$_{3}$) can account for the excess in the polarisation of the starlight from the observed spectrum towards a background star HD 204827, with the high grain alignment degree of $R=0.7$ adopted for oblate grains of very small grains (VSGs) of $27.5\AA\times 5\AA$. To date, no physical mechanism has been demonstrated to produce such a high alignment degree for VSGs. \cite{2014A&A...561A..82S} demonstrated that a combination of silicates, carbon, and graphite grains can self-consistently reproduce both the extinction curve and the polarisation spectrum from UV to IR wavelength towards the HD 37903 star (Serkowski-like spectrum). However, they assumed that grains are aligned by the Davis-Greenstein paramagnetic relaxation mechanism and have perfect internal alignment of the grain axis of major inertia with the grain angular momentum. They also assumed that carbon grains with iron inclusions are aligned, which is challenged by the fact that even though the iron depletion in the carbon-rich circumstellar envelope (CSE) of the asymptotic giant branch star IRC+10\,216 is higher than in diffuse gas \citep[$-2.24$ vs. $-2.00$; ][]{mauron2010, whittet2003}, magnetic alignment has not been confirmed in the CSE. In addition, \citet{chiar2006} showed that the upper limit on the polarisation of the 3.4 $\mu$m aliphatic CH line puts severe limitations on the alignment efficiency of carbon grains.

Recently, \cite{2023ApJ...948...55H} (hereafter HD23) proposed a new composition for dust grains (\texttt{astrodust}), a mixture of silicate and hydrocarbon material. The authors demonstrated that this new type of grain composition can be found in accounts of the standard extinction, emission, and polarisation in the diffuse interstellar medium (ISM). The best-fit alignment function in \cite{2023ApJ...948...55H} corresponds to the negligible alignment of very small grains below $\sim 0.01\,\mu$m and the perfect alignment of large grains above $\sim 0.1\,\mu$m. However, whether the \texttt{astrodust} model with such an alignment function can reproduce the anomalous sightlines with super-Serkowski spectra has not yet been studied. In the first paper in the Grain alignment and Dust Evolution Physics with Polarisation (GRADE-POL) series (\citealt{2025A&A...703A.192T}), we showed that \texttt{astrodust} could nicely reproduce observations ($p_{\rm ext}/A_{\rm V}$ vs. $A_{\rm V}$ for starlight polarisation and $p_{\rm em}$ vs. $A_{\rm V}$ for polarised thermal dust) of the starless core. Within this project, we also presented a numerical model, \texttt{DustPOL\_py}\footnote{\href{https://github.com/lengoctram/DustPOL_py}{https://github.com/lengoctram/DustPOL\_py}}.

The goal of this paper is to test the proposed mechanisms for the excess UV polarisation, including enhanced DG \citep{2014ApJ...790....6H} or enhanced RAT alignment by a stronger radiation field \citep{Hoang.2017SNe} through fitting the theoretical polarisation models to observational data. We first improve our \texttt{DustPOL\_py} tool for more accurate modelling of starlight polarisation and extinction curves for aligned grains. We then evaluated our model predictions against two datasets that ranged from IR to UV, including polarisation spectrum and extinction curve towards HD 37903, HD 161056, HD 204827, and HD 30614, showing both Serkowski-like and super-Serkowski-like spectra. The extinction curves towards these stars (except for HD 161056, because no extinction curve has been measured towards this line of sight) exhibit the $2175\AA$ feature, which prompts us to integrate \texttt{astrodust} with polycyclic aromatic hydrocarbons (PAHs). The locations of these targets are depicted in Figure \ref{fig:Wuppe_targets}.

This paper is structured as follows. We summarise the update in our \texttt{DustPOL\_py} model and the modelled spectrum of the polarisation and extinction curve in Section \ref{sec:modelling}. The fitting of our models to the observations is described in Section \ref{sec:obs_model}. Discussions of the results are given in Section \ref{sec:discussion}. A short summary of our work is given in Section \ref{sec:summary}.

\section{Numerical modelling} \label{sec:modelling}
\subsection{Grain sizes to be aligned by RATs}
The principle of grain alignment by radiative torques (RATs) includes the fundamental effects: spinning the grains up to suprathermal or spinning the grains down to thermal rotations, and the grains precessing around the radiation direction (i.e. radiative precession). Spinning grains acquired magnetic moment through Barnett effect, which allows them to precess around the ambient magnetic fields usually faster than the radiative precession and become aligned with B-fields (see e.g. \citealt{2015ARA&A..53..501A, 2015psps.book...81L, 2022FrASS...9.3927T,2025ApJ...994..115H} for details). The governing formulae used in our model are summarised in Section 2.3 in the first article of this series (\citealt{2025A&A...703A.192T}). In this section, we recall the most important factor for RATs, which is the minimum size of the grains that can be effectively aligned with the magnetic field by RATs, also known as the alignment size $a_{\rm align}$. This alignment size is determined by the requirement for the grain suprathermal rotation, which is given by \citep{2021ApJ...908..218H}
\begin{equation}
    \begin{split}
        a^{\rm RAT}_{\rm min} \equiv a_{\rm align} &= 0.055\,{\rm \mu m} \times U^{-2/7} \left(\frac{\rho}{3\,\rm g\,cm^{-3}}\right)^{-1/7}\left(\frac{\gamma}{0.1}\right)^{-2/7} \\
        & \times \left(\frac{n_{\rm H}}{10^{3}\,\rm cm^{-3}}\right)^{2/7} \left(\frac{T_{\rm gas}}{10\,\rm K} \right)^{2/7}\left(\frac{\bar{\lambda}}{1.2\,\rm \mu m}\right)^{4/7}(1+F_{\rm IR})^{2/7}
    \end{split}
\end{equation}
where $\rho$ is the mass density of the dust grain, $\gamma$ is the anisotropic degree of the radiation field (we adopt $\gamma=0.1$ as in \citealt{1996ApJ...470..551D}), $n_{\rm H}$ and $T_{\rm gas}$ are the gas number density and temperature, and $F_{\rm IR}=0.038\times (a/0.1\,\rm \mu m)^{-1}U^{2/3}(n_{\rm H}/10^{3}\,\rm cm^{-3})^{-1}(T_{\rm gas}/10\,\rm K)^{-1/2}$ is the damping ratio of IR to collision. If $F_{\rm IR}\gg 1$, $a_{\rm align}$ is independent of $n_{\rm H}$ and $T_{\rm gas}$ and depends mainly only on $U$.

For the typical interstellar medium radiation field (ISRF) with the Lyman limit, $a_{\rm align}\simeq 0.05\,\mu$m. Differing the physical parameters will alter the value of $a_{\rm align}$. As an example, for more intense radiation scaling from the ISRF or where EUV photons are allowed to propagate, $a_{\rm align}$ becomes smaller.

\subsection{Alignment degree for grains by RATs}
The alignment size $a_{\rm align}$ indicates the minimum size of aligned grains by RATs, but does not describe how efficient grain alignment is. To fully describe the size-dependent alignment of dust grains, it requires the second important parameter, which measures the alignment degree of the grain axis of major inertia ($\boldsymbol{a}_{1})$ with the ambient magnetic field, which is defined by the Rayleigh reduction factor $R$. The net alignment degree of a non-spherical grain with the ambient magnetic field is achieved through two mechanisms: internal alignment degree efficiency $Q_{\rm X}$, which the alignment between $\boldsymbol{a}_{1}$ and the angular momentum $\boldsymbol{J}$, and the external alignment degree, described by efficiency $Q_{\rm J}$, which characterises the alignment of $\boldsymbol{J}$ with an external axis such as $\boldsymbol{B}$. The net degree of general alignment is given by $f_{\rm max} \equiv R\simeq Q_{\rm X}Q_{\rm J}(1+f_{\rm corr})$ with $f_{\rm corr}=0$ for no correlation. 

For grains aligned by RATs ($a>a_{\rm align}$), the degree of RAT alignment, $R_{\rm RAT}$, depends on the gas properties, the size and magnetic properties of dust, and the radiation field, including intensity and the angle between the radiation direction and the magnetic fields \citep{HoangLaz.2016a} as observationally confirmed by \cite{2011A&A...534A..19A}. The alignment degree can reach $f_{\rm max}= R_{\rm RAT}= 1$ for grains with super-paramagnetic inclusions, likely iron-based because of cosmic abundances, due to the magnetically enhanced RAT mechanism combined with collisional excitations (i.e. MRAT; \citealt{HoangLaz.2016a,2025ApJ...994..115H}). For stronger radiation fields, the maximum alignment by MRAT can be reduced due to the RAT trapping effect \citep{2025ApJ...994..115H}. For an ordinary paramagnetic grain with a typical ISRF, $f_{\rm max}\sim 0.5$ (see \citealt{2021ApJ...913...63H}). In this work, we leave $f_{\rm max}$ as a free parameter.

\subsection{Alignment degree of paramagnetic alignment for small grains} \label{sec:align_degree}
This section describes the alignment of grains with $a<a_{\rm align}$, using the DG alignment, which is the elaboration compared to the previous version of our model (\citealt{2025A&A...703A.192T}). Regarding the internal alignment (\citealt{1979ApJ...231..404P}) of the angular momentum $\boldsymbol{J}$ with the principal axis of the grain, $\boldsymbol{a}_{1}$, \cite{1997ApJ...484..230L} demonstrated that, under thermal equilibrium (where the angle $\theta$ between $\boldsymbol{J}$ and $\boldsymbol{a}_{1}$ follows a Boltzmann distribution), the internal alignment degree is given by 
\begin{equation}
        Q_{\rm X} = \frac{3}{2}\left[\langle \cos^{2} \theta \rangle - \frac{1}{3}\right] = \frac{3}{2\eta^{2}} \left[\frac{\int_{0}^{\eta}t^{2}e^{(t^{2}-\eta^{2})}dt}{\int_{0}^{\eta}e^{(t^{2}-\eta^{2})}dt}\right] - \frac{1}{2},
\end{equation}
where 
\begin{equation}
    \eta^{2}=\frac{(h-1)}{2}\frac{T_{\rm gas}}{T_{\rm dust}}\frac{J^{2}}{J^{2}_{\rm th}}
\end{equation}
is the suprathermality factor with $J^{2}_{\rm th}=k_{\rm B}T_{\rm gas}I$ the thermal angular momentum with $I$ the inertia momentum, $h=2/(1+s^{2})$ for the $s=a/b<1$ ratio of the semi-minor to semi-major axes.

The exact angular momentum of the grain $J^{2}$ of small grains depends on various other damping and excitation processes, including gas collisions, infrared emission, plasma drag, and ion-grain collisions. In the case of purely gas collisions, the rms grain angular momentum, $\langle J^{2}\rangle^{1/2}$, is calculated by the Maxwellian angular momentum given by (\citealt{1997ApJ...484..230L})
\begin{equation} \label{eq:J_Max}
    \frac{J_{\rm Max}}{J_{\rm th}} = \left\{ \left[ 1 + \frac{s^{2}}{2} \right]\left[1+\frac{T_{\rm dust}}{T_{\rm gas}}\right] \right\}^{1/2}.
\end{equation}

To account for realistic situations, a dimensionless parameter, $\alpha_{\rm DG}$, is introduced and defined as $\alpha_{\rm DG}=\langle J^{2}\rangle^{1/2}/J_{\rm Max}$, where $\alpha_{\rm DG}=1$ for the case where the gas damping is dominant and $\alpha_{\rm DG}<1$ when the infrared damping and other interaction processes are taken into account \citep{1998ApJ...508..157D,HoangLaz.2016b}. Based on these constraints, we calculate the value of the internal alignment degree, $Q_{\rm X}(s,\frac{T_{\rm gas}}{T_{\rm dust}}, J=\alpha_{\rm DG} J_{\rm Max})$. We note that the damping of IR is important for small grains and a lower $\alpha_{\rm DG}$ corresponds to more importance in IR damping. Moreover, because we used four different lines of sight (LOSs) with different conditions, we varied this parameter in this work as a free parameter.

For the external alignment between the angular momentum $\boldsymbol{J}$ and the alignment axis (e.g. magnetic field), using the Fokker-Planck equation approach, \cite{1967ApJ...147..943J} derived the external alignment efficiency as
\begin{equation}
    Q_{\rm J} = \frac{3}{2} \times \left\{
    \begin{array}{l l}
        -\frac{1}{3}+\frac{1}{x}\left[\left(\frac{1+x}{x}\right)^{1/2}{\rm arcsinh}\sqrt{x}-1\right] & \quad {\rm ~ for ~} x>0 \\
        -\frac{1}{3}+\frac{1}{x}\left[\left(\frac{1+x}{-x}\right)^{1/2}\arcsin \sqrt{-x}-1 \right]  & \quad {\rm ~ for ~} x<0 \\
        0 & \quad {\rm ~ for ~} x= 0
    \end{array} \right.\label{eq:QJ_JS67}
\end{equation}
where $x=\left(\frac{\delta_{\rm m}}{1+\delta_{\rm m}}\times \frac{T_{\rm dust}-T_{\rm gas}}{T_{\rm gas}}\right)$, and $\delta_{\rm m}=\tau_{\rm DG}^{-1}/\tau_{\rm gas}^{-1}$ measures the competition between grain alignment by paramagnetic relaxation and disalignment by random gas collisions. We note that the above formula is also derived for thermally rotating grains by assuming that grain rotation is only influenced by the gas collisions, disregarding other interaction processes such as infrared emission and plasma drag. Thus, Equation \ref{eq:QJ_JS67} provided the upper limit for the DG mechanism, and the realistic degree of DG alignment is much lower due to the grain's subthermal rotation (see \citealt{HoangLazMartin.2014} for more details).

The DG mechanism is more effective for $\delta_{\rm m} > 1$. In general, the relation of $\delta_{\rm m}$ to $a$ is given by (see Section \ref{app:grain_phys} for more details)
\begin{equation} \label{eq:delta_m}
    \begin{split}
        \delta_{\rm m} &= \frac{\tau_{\rm gas}}{\tau_{\rm DG}} = \frac{\sqrt{\pi}K(\omega)B^{2}}{1.2\sqrt{2k_{\rm B}m_{\rm H}}n_{\rm H}\sqrt{T_{\rm gas}}a\Gamma_{\parallel}} \\
        &\simeq 4.56N_{\rm cl}\phi_{\rm sp}\Gamma^{-1}_{\parallel}\left(\frac{a}{0.1\rm \,\mu m}\right)^{-1} \left(\frac{n_{\rm H}}{10^{4}\,\rm cm^{-3}}\right)^{-1}\left(\frac{T_{\rm gas}}{10\,\rm K}\right)^{-0.5} \\
        &\times \left(\frac{T_{\rm dust}}{10\,\rm K}\right)^{-1}
        \left(\frac{B}{20\,\rm \mu G}\right)^{2}\exp{\left(\frac{0.011{\rm K} \times N_{\rm cl}}{T_{\rm dust}}\right)}\left[1+\left(\frac{\omega\tau_{\rm sp}}{2}\right)^{2}\right]^{-2},
    \end{split}
\end{equation}
where $N_{\rm cl}$ is the number of iron cluster and $\phi_{\rm sp}$ is the volume filling factor of the iron cluster inside the grain. For a paramagnetic grain, $N_{\rm cl}=1$ and $\phi_{\rm sp}=1/7$, while $N_{\rm cl}$ is much higher for super-paramagnetic grains. $\Gamma_{\parallel}$ is the geometrical factor of the grain and $\tau_{\rm sp}$ is the remagnetisation time, the frequency $\omega \simeq \frac{(h-1)}{2} J/I_{\parallel}$, approximately to the angular velocity. $K(\omega)$ characterises the magnetic susceptibility of a grain rotating at an angular velocity $\omega$. The degree of net alignment is 
\begin{equation} \label{eq:fmin}
    f_{\rm min}\equiv R_{\rm DG}(a)=Q_{\rm X}Q_{\rm J}.
\end{equation}
Figure \ref{fig:R_DG} illustrates the degree of alignment of DG with grain size. As in $\delta_{\rm m} \sim a^{-1}$, larger grains tend to have a lower degree of alignment. However, in the presence of a strong magnetic field, a higher number of large grains can be aligned, resulting in a uniform alignment degree. It is important to note that a higher $T_{\rm gas}/T_{\rm dust}$ ratio also increases alignment efficiency, as the dissipation of energy to heat becomes more effective, causing $\boldsymbol{J}$ to align more closely with $\boldsymbol{B}$.

In essence, the degree of alignment of the DG mechanism is affected by four factors: $T_{\rm gas}/T_{\rm dust}$, the rotational parameter $\alpha_{\rm DG}$, the magnetic susceptibility $\chi$, and $B$. In this work, to reduce the degeneracy, we adopt the relation of the magnetic field strength on the line of sight and the volume density of the gas as $|B|_{\rm los}=10\times (n_{\rm H}/300{\rm cm}^{-3})^{0.65}\,\mu$G for $n_{\rm H}\geq 300\,\rm cm^{-3}$, and 10$\,\mu$G otherwise following \cite{2010ApJ...725..466C}. The total magnetic field strength is then $\langle |B| \rangle = 2|B|_{\rm los}$. Furthermore, the equilibrium $T_{\rm dust}$, estimated from the radiation intensity, is used.

\subsection{Alignment degree of paramagnetic resonant alignment for very small grains}
\begin{figure}[!ht]
    \centering
    \includegraphics[width=1.0\linewidth]{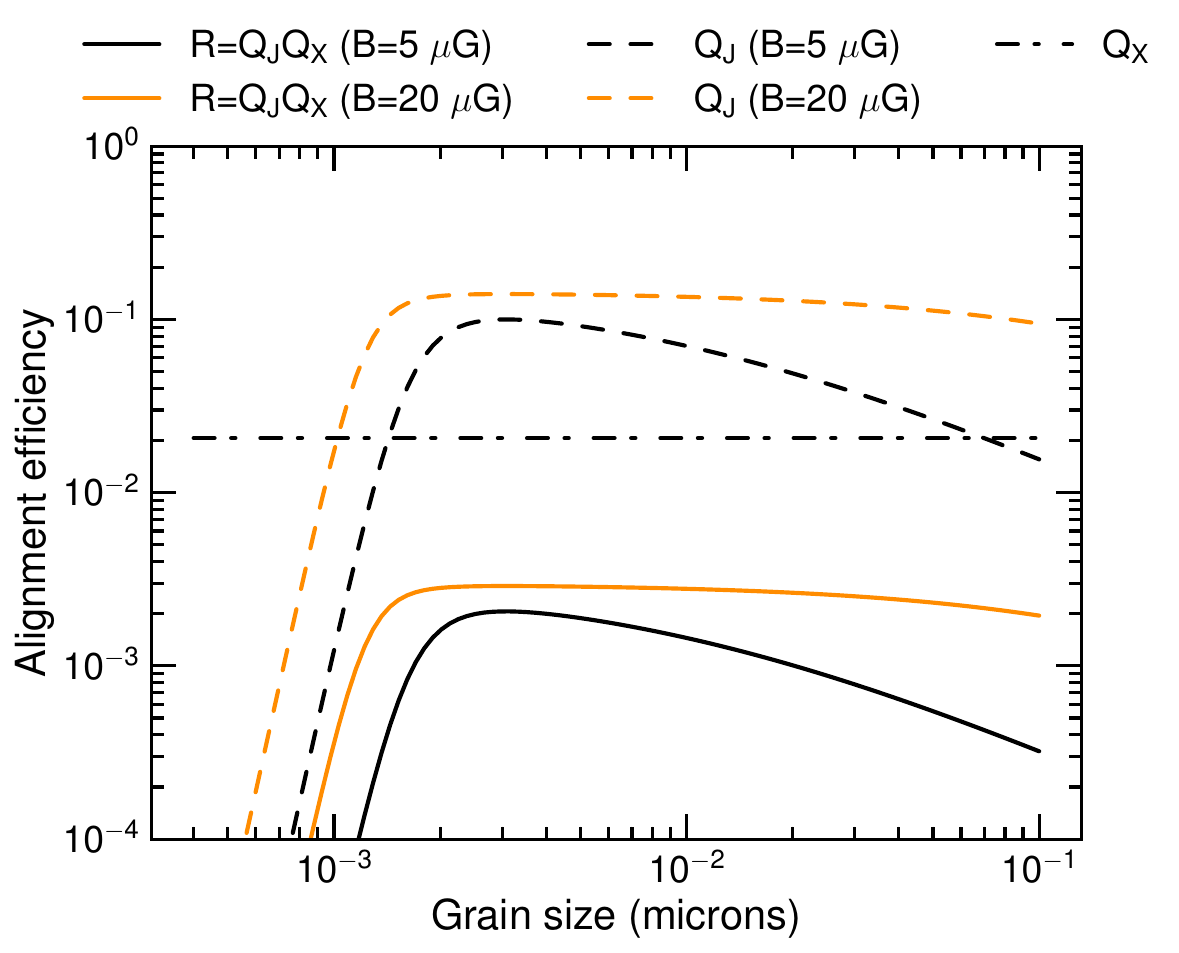}
    \caption{Degree of paramagnetic alignment with the grain size for $n_{\rm H}=10^{3}\,\rm cm^{-3}$. The alignment degree decreases with larger grain sizes; however, a stronger magnetic field can enhance the alignment for such large grains (orange lines). We use $U=1$ for this plot.}
    \label{fig:R_DG}
\end{figure}

As illustrated in Figure \ref{fig:R_DG}, the DG alignmnet becomes inefficient, for example, those less than $10^{-3}\,\mu$m. This section describes the efficient alignment of these very small grains. Taking into account the splitting in the rotational energy, \cite{2000ApJ...536L..15L} found that paramagentic relaxation can occur resonantly when the grains are rotating with the ambient magnetic field. This alignment is called resonant paramagnetic relaxation. Therefore, the efficiency of the external alignment $Q_{\rm J}$ is altered. Following \cite{2014ApJ...790....6H}, $\delta^{\rm res}_{\rm m}$ is calculated using Equation \ref{eq:delta_m} by replacing the $K(\omega)$ term with
\begin{equation}
    K(\omega) = \frac{\chi(0)\tau_{2}}{1+\gamma^{2}g^{2}_{e}\tau_{1}\tau_{2}H^{2}_{1}\sin^{2}\theta}
\end{equation}
with
\begin{equation}
    \begin{split}
        \gamma^{2}g^{2}_{e}\tau_{1}\tau_{2}H^{2}_{1}\sin^{2}\theta &= 8\times\left(\frac{\tau_{1}}{10^{6}\,\rm s}\right)\left(\frac{\tau_{2}}{2\times 10^{-9}\,\rm s}\right)\left(\frac{H_{1}}{5\,\rm \mu G}\right)^{2}\left(\frac{\sin^{2}\theta}{2/3}\right) \\
        \tau_{1} &= 10^{-6} \times \left(\frac{77}{T_{\rm d}}\right)^{m+1}\left(\frac{T_{\rm d}}{T_{l}}\right)^{m}e^{T_{\rm d}/T_{l}}m! \zeta(m) ~~~{\rm s} \\
        T_{l} &= 63 \times \left(\frac{10^{-7}\,\rm cm}{a}\right) ~~~ {\rm K},
    \end{split}
\end{equation}
where $\tau_{1}$ is the spin-lattice relaxation time, $T_{l}$ is the lowest grain vibrating temperature, and $\zeta(m)$ is the Riemann zeta function. In this work, we take $m=6$, $H_{1}\equiv B$, and $\tau_{2}\equiv \tau_{\rm sp}$. One can see that the timescale ratio is $\delta_{\rm m}=\max(\delta^{\rm DG}_{\rm m},\, \delta^{\rm res}_{\rm m})$.

\begin{figure}[!ht]
    \centering
    \includegraphics[width=0.95\linewidth]{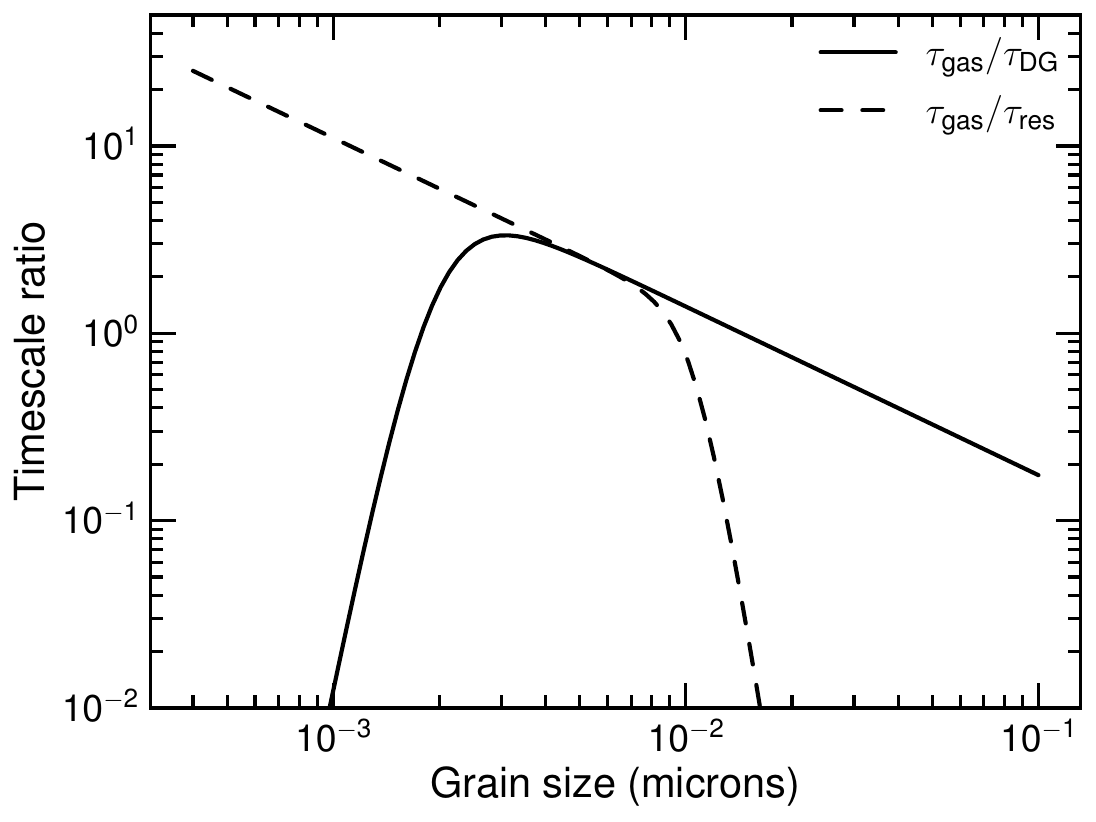}
    \caption{Comparison on the timescale between the paramagnetic alignment for small grains (solid line) and very small grains (dashed line).}
    \label{fig:R_res}
\end{figure}

Figure \ref{fig:R_res} illustrates the relationships between $\delta^{\rm DG}_{\rm m}$ and $\delta^{\rm res}_{\rm m}$ with grain size. As anticipated, $\delta^{\rm res}_{\rm m}>\delta^{\rm DG}_{\rm m}$ towards the smallest grain size. The alignment efficiency is then calculated as in Equation \ref{eq:fmin}.

\begin{table*}[]
    \centering
    \caption{Basic information of the targets used in our work.}
    \begin{tabular}{l|c c c c c c}
         \hline
         \hline
         Target & RA & Dec & $R_{\rm V}$ & \multicolumn{3}{c}{Serkowski fitting parameters$^{a}$}  \\
         {}     & {} & {}  & {}         & $p_{\rm max}$ (\%) & $\lambda_{\rm max}$ ($\mu$m) & $K$ \\
         \hline
         HD 37903 & 05h41m38.388s & -02d15m32.476s & 4.11$^{b}$ & 1.98 & 0.692 & 1.160 \\
         HD 161056& 17h43m47.020s & -07d04m46.581s & {}   & 4.02 & 0.591 & 0.990 \\
         HD 30614 & 04h54m03.011s & +66d20m33.641s & 3.01$^{c}$ & 1.61 & 0.514 & 0.863 \\
         HD 204827& 21h28m57.761s & +58d44m23.238s & 2.45$^{d}$ & 5.67 & 0.413 & 0.696 \\
         \hline
         \multicolumn{7}{l}{$^{a}$ taken from \cite{1995ApJ...445..947C}} \\
         \multicolumn{7}{l}{$^{b}$ taken from \cite{2009ApJ...705.1320G}} \\
         \multicolumn{7}{l}{$^{c}$ taken from \cite{2021ApJS..257...63Z}} \\
         \multicolumn{7}{l}{$^{d}$ taken from \cite{2005AJ....130.1127F}}
    \end{tabular}
    \label{tab:target_info}
\end{table*}
\begin{table*}
    \centering
    \caption{Fixed parameters in our model}
    \begin{tabular}{c|c|c}
        \hline 
        \hline
        Parameters & Value & Note  \\
        \hline
        $b/a$ & 1.4 & Grain's aspect ratio (Oblate grain)\\
        $a_{\rm min}$ & 3.1\AA & Lower limit in the size distribution of grain\\
        $a_{\rm max}$ & $0.5\,\mu$m & Upper limit in the size distribution of grain$^{*}$\\
        $\psi$ & $90^\circ$ & Inclination angle of Bfield (on the plane-of-sky)\\
        $a^{\rm PAH}_{0,1}$ & 4\AA & Peak 1 for PAHs size distribution\\
        $a^{\rm PAH}_{0,2}$ & 30\AA & Peak 2 for PAHs size distribution\\
        $\sigma_{\rm PAH}$ & 0.4 & Widths of log-normal size distribution for PAHs\\
        $a_{0,\rm Ad}$ & 63.8\AA & Peak of log-normal size distribution for astrodust\\
        $\sigma_{\rm Ad}$ & 0.353 & Width of the log-normal size distribution for astrodust\\
        $B_{\rm Ad}$ & $3.3\times 10^{-10}$& A parameter for log-normal size distribution for astrodust\\
        $A_{0}$ & $2.97\times 10^{-5}$ & A parameter for non-log-normal size distribution for astrodust\\
        $A_{4}$ & $7.96\times 10^{-3}$ & A parameter for non-log-normal size distribution for astrodust\\
        $A_{5}$ & $-1.68\times 10^{-3}$ & A parameter for non-log-normal size distribution for astrodust\\
        \hline
    \end{tabular}
    \tablefoot{$^{*}$ We made a series of test by varying $a_{\rm max}$, but best values are slightly scattered around $0.5\,\mu$m. Thus, we fixed this value to reduce the number of free parameter.}
    \label{tab:params_fixed}
\end{table*}

\begin{table*}
    \centering
    \caption{Derived parameters from our model to Serkowski and super-Serkowski spectra, according to the highest likelihood}
    \begin{tabular}{c|l|c c c c c c c c c c}
        \hline
        \hline
        {} & \multirow{4}{4em}{Target} & \multicolumn{9}{c}{{\large \bf Cold Neutral Medium}} \\
        {} & {} & \multicolumn{9}{c}{$n_{\rm H}=30\,\rm cm^{-3}$, $T_{\rm gas}=100\,$K, $\bar{\lambda}=1.2\,\mu$m, $\gamma=0.1$} \\
        \cline{3-12}
        {} & {} & \multicolumn{9}{c}{{\bfseries\itshape\boldmath RAT ($a_{\rm align}=0.05\,\mu\mathrm{m}$ fixed) + DG alignments}} \\
        {} & {} & $f_{\rm max}$ & $U$ & $\alpha_{\rm DG}$ & $A_{1}$ & $A_{2}$ & $A_{3}$ & $B_{1}$ & $B_{2}$ & $N_{\rm H}$ & $\chi^{2}_{\rm p_{\rm ext}}/N_{\rm points}$\\

         \cline{2-12}
         \multirow{10}{*}{\rotatebox[origin=c]{90}{super-Serkowski}} & HD 30614 & 0.35 & - & 0.99 & -0.10 & -1.59 & 0.20 & 6.43(-7) & 2.95(-9) & - & 9.73\\
         & HD 204827 & 0.31 & - & 0.97 & -0.11 & -1.61 & 0.20 & 4.33(-8) & 1.28(-9) & 1.74(22) & 1.69 \\

        \cline{2-12}
        {} & {} & \multicolumn{10}{c}{{\bfseries\itshape\boldmath Only RAT ($f_{\rm min}=0$)}} \\
        {} & {} & $f_{\rm max}$ & $U$ & $\alpha_{\rm DG}$ & $A_{1}$ & $A_{2}$ & $A_{3}$ & $B_{1}$ & $B_{2}$ & $N_{\rm H}$ \\

         \cline{2-12}
         {} & HD 30614 & 0.49 & 9.67(4) & - & -2.73 & -0.97 & 0.17 & 1.32(-6) & 3.76(-10) & - & 6.89\\
         & HD 204827 & 0.45 & 0.85(2) & - & -0.20 & -1.63 & 0.21 & 2.46(-8) & 2.12(-9) & 1.12(22) & 1.46 \\
        
        \cline{2-12}
        {} & {} & \multicolumn{9}{c}{{\bfseries\itshape DG + RAT alignments}} \\
        {} & {} & $f_{\rm max}$ & $U$ & $\alpha_{\rm DG}$ & $A_{1}$ & $A_{2}$ & $A_{3}$ & $B_{1}$ & $B_{2}$ & $N_{\rm H}$ \\

        \cline{2-12}
        {} & HD 30614 & 0.46 & 9.78(4) & 0.84 & -2.64 & -0.99 & 0.17 & 1.21(-6) & 1.43(-9) & - & 6.91\\
        & HD 204827 & 0.54 & 2.11(3) & 0.17 & -1.64 & -1.29 & 0.19 & 2.71(-9) & 1.00(-9) & 2.10(22) & 1.40 \\
        \hline
    \end{tabular}
    
    \vspace{3mm}
    
    \begin{tabular}{c|l|c c c c c c c c c c}
        \hline
        {} & \multirow{4}{2em}{} & \multicolumn{9}{c}{{\large \bf Interstellar super-bubble}} \\
        {} & {} & \multicolumn{9}{c}{$T_{\rm gas}=10000\,$K, $\bar{\lambda}=0.36\,\mu$m, $\gamma=1$, bubble's radius=$30\,$pc} \\
        {} & {} & \multicolumn{9}{c}{$^{*}$For HD 30614: $n_{\rm H}=2\,\rm cm^{-3}$ -- For HD 204827: $n_{\rm H}=30\,\rm cm^{-3}$} \\

        \cline{3-12}
        {} & {Target} & \multicolumn{9}{c}{{\bfseries\itshape\boldmath Only RAT ($f_{\rm min}=0$)}} \\
        {} & {} & $f_{\rm max}$ & $U$ & $\alpha_{\rm DG}$ & $A_{1}$ & $A_{2}$ & $A_{3}$ & $B_{1}$ & $B_{2}$ & $N_{\rm H}$ & $\chi^{2}_{\rm p_{\rm ext}}/N_{\rm points}$\\

         \cline{2-12}
          \multirow{6}{*}{\rotatebox[origin=c]{90}{super-Serkowski}} & HD 30614 & 0.54 & 4.09(2) & - & -2.84 & -0.94 & 0.17 & 1.27(-6) & 1.75(-10) & - & 6.87\\
          & HD 204827 & 0.48 & 0.27(2) & - & -1.93 & -1.22 & 0.18 & 4.57(-8) & 7.61(-10) & 2.57(22) & 1.41\\
        
        \cline{2-12}
        {} & {} & \multicolumn{9}{c}{{\bfseries\itshape DG + RAT alignments}} \\
        {} & {} & $f_{\rm max}$ & $U$ & $\alpha_{\rm DG}$ & $A_{1}$ & $A_{2}$ & $A_{3}$ & $B_{1}$ & $B_{2}$ & $N_{\rm H}$ \\

        \cline{2-12}
        {} & HD 30614 & 0.46 & 1.23(2) & 0.19 & -3.07 & -0.88 & 0.16 & 1.53(-6) & 1.14(-10) & - & 6.43\\
        & HD 204827 & 0.43 & 0.19(2) & 0.19 & -2.82 & -0.97 & 0.16 & 2.46(-7) & 1.49(-10) & 2.18(22) & 1.27\\
        \hline
    \end{tabular}
    
    \vspace{3mm}
    
    \begin{tabular}{c|l|c c c c c c c c c c}
        \hline
        {} & \multirow{4}{4em}{} & \multicolumn{9}{c}{{\large \bf Cold Neutral Medium}} \\
        {} & {} & \multicolumn{9}{c}{$n_{\rm H}=30\,\rm cm^{-3}$, $T_{\rm gas}=100\,$K, $\bar{\lambda}=1.2\,\mu$m, $\gamma=0.1$} \\

        \cline{3-12}
        {} & {Target} & \multicolumn{10}{c}{{\bfseries\itshape\boldmath RAT ($a_{\rm align}=0.05\,\mu$m fixed) + DG alignment}} \\
        {} & {} & $f_{\rm max}$ & $U$ & $\alpha_{\rm DG}$ & $A_{1}$ & $A_{2}$ & $A_{3}$ & $B_{1}$ & $B_{2}$ & $N_{\rm H}$ & $\chi^{2}_{\rm p_{\rm ext}}/N_{\rm points}$\\

        \cline{2-12}
        \multirow{6}{*}{\rotatebox[origin=c]{90}{Serkowski}} & HD 37903 & 0.85 & - & 1.84(-2) & -4.57 & -0.66 & 0.16 & 1.54(-7) & 5.97(-10) & - & 0.65\\
         & HD 161056 & 0.49 & - & 5.05(-2) & -4.62 & -0.59 & 0.14 & - & - & 5.00(22) & 1.39\\

        \cline{2-12}
        {} & {} & \multicolumn{9}{c}{{\bfseries\itshape\boldmath Only RAT ($f_{\rm min}=0$)}} \\
        {} & {} & $f_{\rm max}$ & $U$ & $\alpha_{\rm DG}$ & $A_{1}$ & $A_{2}$ & $A_{3}$ & $B_{1}$ & $B_{2}$ & $N_{\rm H}$ \\

         \cline{2-12}
         {} & HD 37903 & 0.85 & 0.51 & - & -4.68 & -0.60 & 0.15 & 2.09(-7) & 3.21(-10) & - & 0.57\\
         & HD 161056 & 0.37 & 1.50 & - & -4.61 & -0.57 & 0.14 & - & - & 2.89(22) & 1.44 \\

        \hline
    \end{tabular}
    \tablefoot{\\
    - $a(b) = a\times 10^{b}$ \\ 
    - The best value of $U$ is derived from the best fitting-parameter $U_{\rm scale}$. \\ 
    - $^{*}$ The physical reasons for choosing different gas density for HD 30614 and HD 204827 are given in the main text.}
    \label{tab:params_derived}
\end{table*}

\subsection{Size-dependence alignment degree: Connecting paramagnetic and RAT alignment}
To account for the transition of the degree of alignment $R$ from $a<a_{\rm align}$ to $a>a_{\rm align}$, we apply a smooth function similar to that described in 
\cite{2020ApJ...896...44L}, which is known as the alignment function,
\begin{equation}
    f(a) = f_{\rm min} + (f_{\rm max}-f_{\rm min})\left[1-e^{-(0.5a/a_{\rm align})^{3}}\right],\label{eq:falign}
\end{equation}
where $f \rightarrow f_{\rm min}$ for $a\ll a_{\rm align}$ and $f \rightarrow f_{\rm max}$ for $a\gg a_{\rm align}$. We recall that $f_{\rm min}$ is estimated using Equation \ref{eq:fmin} while $f_{\rm max}$ is the maximum alignment degree of grains by RATs. Here, $f_{\rm max}$ is a model parameter.

\subsection{Extinction curve and spectrum of starlight polarisation}
In this section, we improve the estimation of the extinction curve and polarisation spectrum for a population of partially aligned grains, formulated in the \texttt{DustPOL\_py} model. The extinction curve and the degree of starlight polarisation (in $\%$), assuming only \textsc{astrodust} is aligned, are calculated as (see Appendix \ref{sec:cross-sections} for more details)
\begin{equation}
    \begin{split}
        \frac{A_{\lambda}}{N_{\rm H}} &= 1.068\times \sum_{\rm i=Ad,\, PAH} \int_{(a)} \langle C_{\rm ext,i}\rangle \frac{1}{n_{\rm H}}\frac{dn_{i}}{da} da \\
        &= 1.068 \times \sum_{\rm i=Ad,\, PAH} \int_{(a)}\left[C^{\rm ran}_{\rm ext} + f(a)C_{\rm pol}\left(\frac{2}{3}-\sin^{2}\psi\right)\right]\frac{1}{n_{\rm H}}\frac{dn_{i}}{da}da
    \end{split}
\end{equation}
and
\begin{equation} \label{eq:pext}
    \frac{P_{\rm ext}}{N_{\rm H}} = 100 \times \int_{(a)} f(a)C_{\rm pol, Ad}\sin^{2}\psi\frac{1}{n_{\rm H}}\frac{dn_{\rm Ad}}{da}da
\end{equation}
where $dn/da$ is the grain size distribution, $\psi$ is the inclination angle of the magnetic field with respect to the LOS, and $C^{\rm ran}_{\rm ext}$ is the extinction cross-section for the randomly oriented grains. 

In this work, we consider combinations of \textsc{astrodust} and PAHs with the same form of grain size distributions as in HD23 
\begin{equation}
    \frac{1}{n_{\rm H}} \frac{dn_{\rm Ad}}{da} = \frac{B_{\rm Ad}}{a}\exp\left(-\frac{[\ln(a/a_{0,Ad})]^{2}}{2\sigma^{2}_{\rm Ad}}\right) + \frac{A_{0}}{a}\exp\left(\sum_{i=1}^{5}A_{i}[\ln(a/\AA)]^{i}\right)
\end{equation}
\begin{equation}
    \frac{1}{n_{\rm H}}\frac{dn_{\rm PAH}}{da} = \sum_{j=1}^{2}\frac{B_{j}}{a}\exp\left(-\frac{[\ln(a/a^{\rm PAH}_{0,j})]^{2}}{2\sigma_{\rm PAH}^{2}}\right).
\end{equation}
Here, $B_{\rm Ad}$, $a_{0,\rm Ad}$, $\sigma_{\rm Ad}$, $A_{0}$, $A_{4}$, $A_{5}$, $a^{\rm PAH}_{0,1}$, $a^{\rm PAH}_{0,2}$, and $\sigma_{\rm PAH}$ are constants, detailed in Table \ref{tab:params_fixed}. In contrast, $A_{1}$, $A_{2}$, $A_{3}$, $B_{1}$, and $B_{2}$ are variable parameters. These parameters determine the shape of the size distribution. The grain size and wavelength dependence of $C_{\rm ext,\mathbf{E\perp a}}$ and $C_{\rm ext, \mathbf{E\parallel a}}$ are taken from \textsc{astrodust} database\footnote{\href{http://arks.princeton.edu/ark:/88435/dsp01qb98mj541}{http://arks.princeton.edu/ark:/88435/dsp01qb98mj541}}. The extinction and absorption cross-sections for PAHs are \cite{2007ApJ...657..810D}. The total-to-selective extinction ratio $R_{\rm V}=A_{\rm V}/(A_{\rm B}-A_{\rm V})$.

Figure \ref{fig:physical_variations} shows an example of how the maximum wavelength ($\lambda_{\rm max}$) and the width ($K$) of the IR to UV polarisation spectrum vary with physical parameters, including the density of the gas volume, the maximum grain size and the radiation intensity. $\lambda_{\rm max}$ is mainly influenced by $a_{\rm max}$, in which smaller grains results in lower value of $\lambda_{\rm max}$. Whereas $K$ is affected by all these parameters. With a fixed $n_{\rm H}$, as $a_{\rm max}$ increases, $K$ shows a negative correlation with $\lambda_{\rm max}$. In contrast, when $a_{\rm max}$ is constant, increasing $n_{\rm H}$ results in a positive correlation between $K$ and $\lambda_{\rm max}$.
        
\section{Applications to observations} \label{sec:obs_model}
In this section, we compare our model predictions to observations for both super-Serkowski and Serkowski lines of sight. For the super-Serkowski, as mentioned in the Introduction, there are three possibilities that we expect to account for the excess in UV polarisation. The first option is that alignment of small grains by the DG mechanism was suggested as the dominant mechanism for excess UV polarisation \citep{HoangLazMartin.2014}. The second option relies only on the ability of the RAT alignment to align smaller grains. For instance, a stronger radiation field can help smaller grains align by RATs, and thus only the RAT alignment by itself can account for the observed super-Serkowski spectrum. The third option is to rely on the EUV radiation field from the ionised ISM, such as the super-bubble. We aim to unravel these three possibilities within two distinct physical conditions: cold neutral medium (CNM) and warm ionised medium (WIM).

\subsection{Interpretations of super-Serkowski spectra with the CNM condition and the scaled ISRF radiation} \label{sec:supSep_CNM}
\subsubsection{Observational targets}
In our target selections, the super-Serkowski sight lines include HD 30614 and HD 204827. The observed polarisation spectra were adopted from \cite{1995ApJ...445..947C}. The extinction curves for HD 30614 were adopted from \cite{2021ApJS..257...63Z}, which is a combination of the FM curve (\citealt{1990ApJS...72..163F}) for $\lambda^{-1}>3.3\,\rm \mu m^{-1}$ and the CCM curve (\citealt{1989ApJ...345..245C}) for $1.1<\lambda^{-1}<3.3\,\rm \mu m^{-1}$. The extinction curve for HD 204827 was reconstructed from \cite{2005AJ....130.1127F}, which is a combination of the FM curve in the UV bands with three anchor points in the optical (3300, 4000, and 5530$\,\AA$) and in the infrared bands (we refer to their Section 3.2 and Table 2 for details).

To constrain the best parameters, we used MCMC sampling with \textsc{emcee} (\citealt{emcee}). The log-likelihood function is defined as
\begin{equation}
    {\rm log-likelihook} = -0.5\times \left(\chi^{2}_{\rm pext} + \chi^{2}_{A_{\lambda}/A_{\rm V}} + \chi^{2}_{R_{\rm V}}\right)
\end{equation}
with 
\begin{equation}
    \begin{split}
        \chi^{2}_{\rm pext} &= \sum\left(\frac{p^{\rm obs}_{\rm ext}-p^{\rm model}_{\rm ext}}{\sigma_{p_{\rm ext}}}\right)^{2}, \\
        \chi^{2}_{A_{\lambda}/A_{\rm V}} &=  \sum \left( \frac{\left(A_{\lambda}/A_{\rm V}\right)^{\rm obs}-\left(A_{\lambda}/A_{\rm V}\right)^{\rm model}}{\sigma_{A_{\lambda}}}\right)^{2}, \\
        \chi^{2}_{R_{\rm V}} & = \sum \left(\frac{R^{\rm obs}_{\rm V}-R^{\rm model}_{\rm V}}{R^{\rm obs}_{\rm V}}\right)^{2}.
    \end{split}
\end{equation}
The fixed parameters used in our model are listed in Table \ref{tab:params_fixed} and the derived values of the free parameters are in Table \ref{tab:params_derived}.

\begin{figure*}
    \centering
    \includegraphics[width=0.98\linewidth]{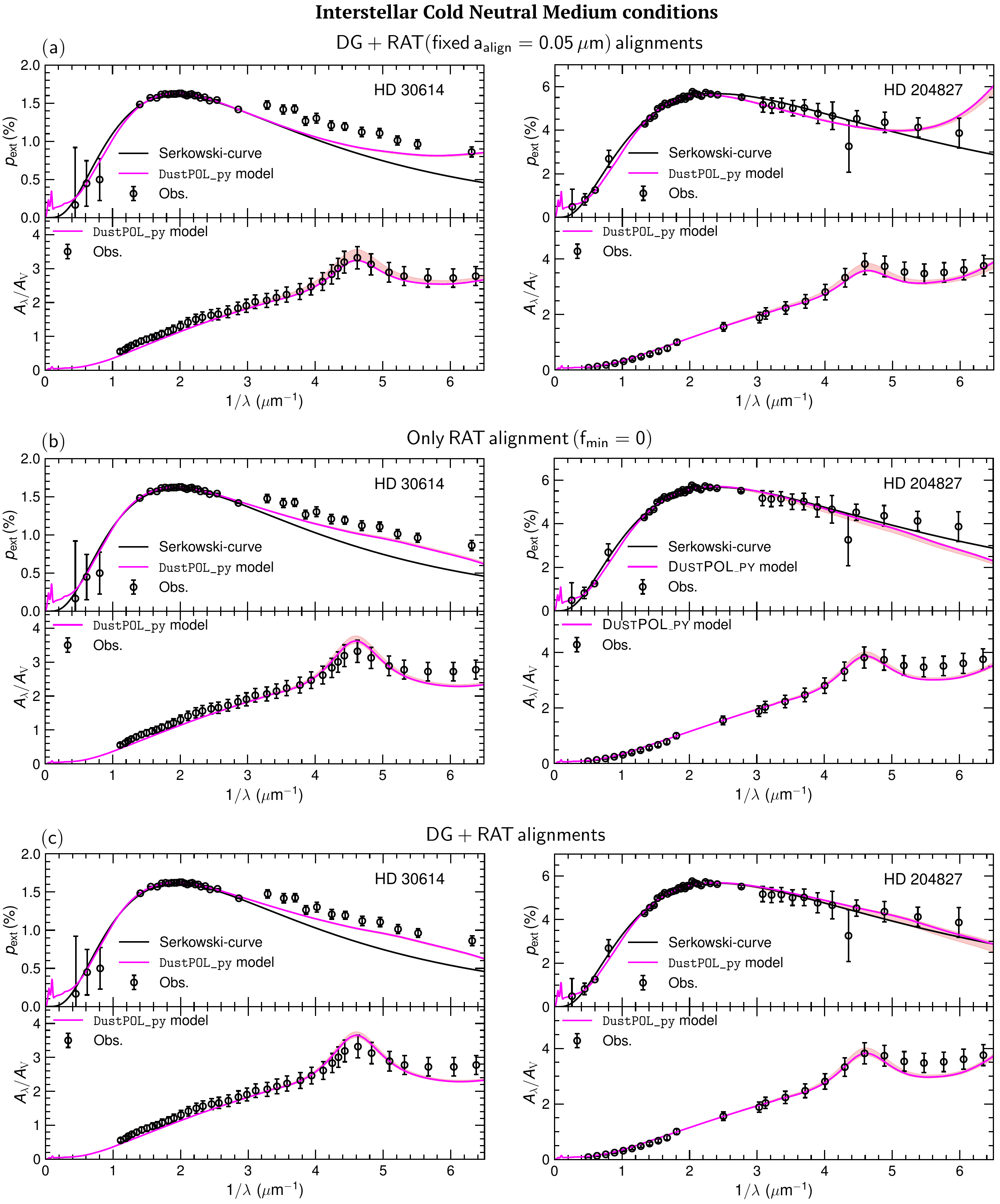}
    \caption{Comparison between our model and the observed super-Serkowski spectra towards HD 30614 (left column) and HD 204827 (right column) with the conditions of the interstellar CNM, $n_{\rm H}=30\,\rm cm^{-3}$, $T_{\rm gas}=100\,$K, and the radiation field scaled from the ISRF. The shaded area is from the top 500 models with the highest posterior probability shown by the solid magenta lines. \textit{Panel (a)}: Combination of RAT alignment with a fixed $a_{\rm align}=0.05\,\mu$m and the DG alignment. \textit{Panel (b)}: Only the RAT alignment with varied $a_{\rm align}$. \textit{Panel (c)}: Combination of the RAT ($a_{\rm align}$ is allowed to vary) and DG alignment. The error for the extinction curve is taken as 10\% of the data. Our models with a joint effect of the RAT and DG alignments self-consistently reproduce the observed data for both super-Serkowski and the extinction curve. However, the radiation strength is extremely higher than the typical ISRF.}
    \label{fig:obs_model_supSep}
\end{figure*}

\begin{figure*}
    \includegraphics[width=1.0\linewidth]{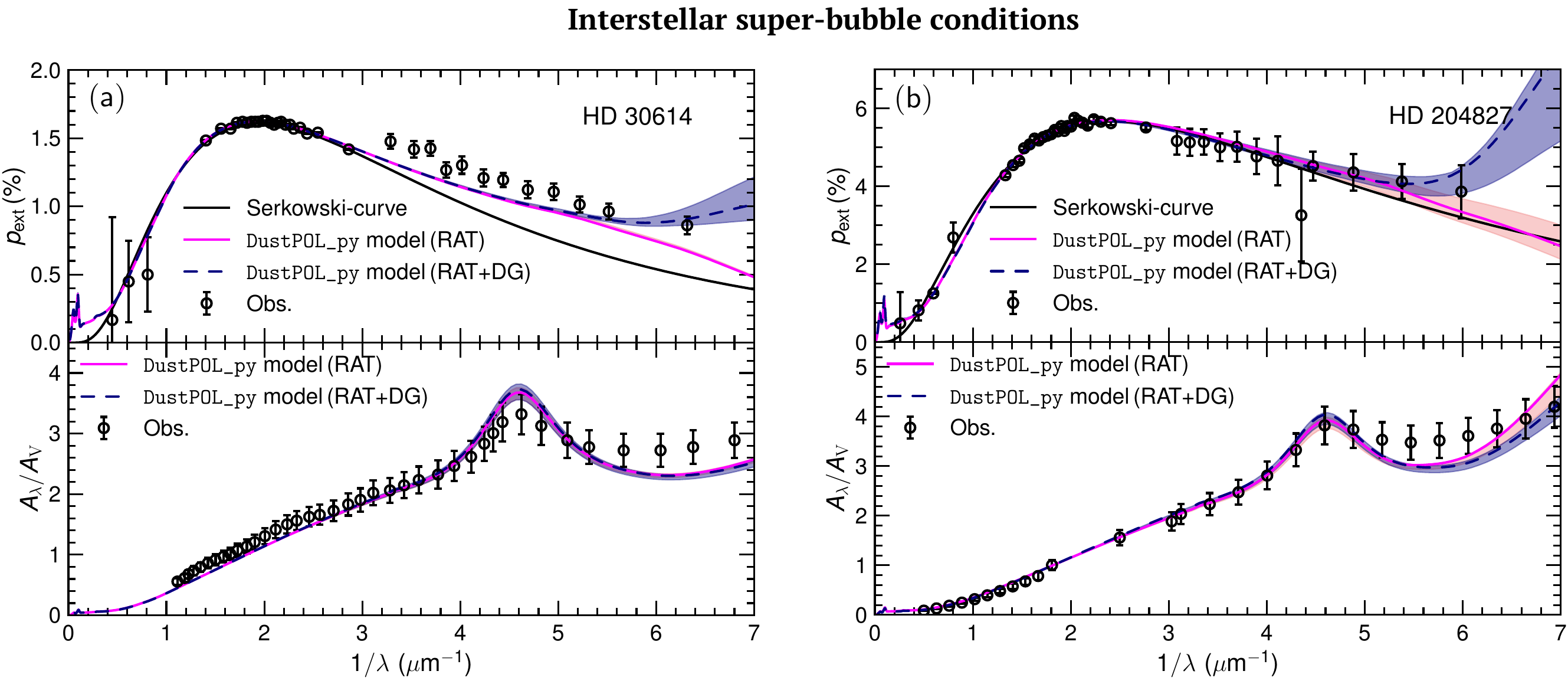}
        \caption{Polarisation spectrum and extinction curve towards super-Serkowski spectra in HD 30614 (\textit{panel a}) and HD 204827 (\textit{panel b}) with models accounting for the grain alignment by RATs and with a combination with the DG alignment for a typical condition of the interstellar super-bubble, $n_{\rm gas}=2\,\rm cm^{-3}$ (HD 30614) and $n_{\rm H}=30\,\rm cm^{-3}$ (HD 204827) and $T_{\rm gas}=10000\,$K. The model nicely reproduces both curves, and these best models are almost identical. Towards the highest wavenumber ($\lambda^{-1}>5.5\,\rm \mu m^{-1}$) the contribution of the DG to the RAT alignment is slightly better compared to the RAT alignment alone. However, because there were no data, the best-fit model shows poor convergence towards the far-UV range.}
        \label{fig:HD30614_EUV}
    \vspace{5mm}
    \centering
    \includegraphics[width=1.0\linewidth]{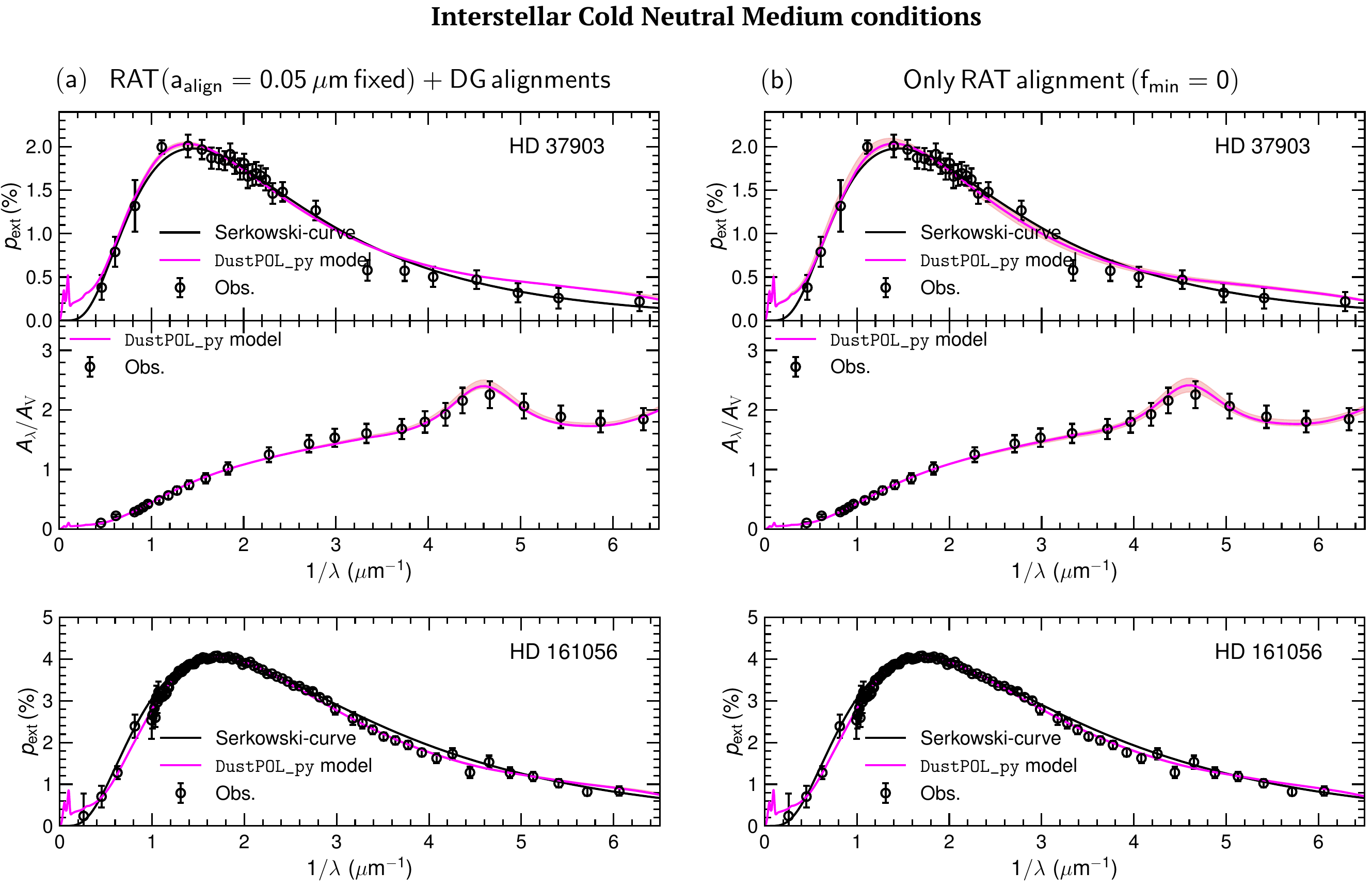}
    \caption{Similar to Figure \ref{fig:obs_model_supSep} but for the targets with Serkowski spectra in HD 37903 and HD 161056. The best model incorporating the DG alignment (panel a) corresponds to the IR damping of $\alpha_{\rm DG} \sim 10^{-2}$, representing an inefficient alignment degree. Only RAT alignment with no DG alignment (panel b) could identically explain the observed data. Both features indicates that the RAT alignment itself can reproduce the Serkowski spectra without the need of the DG alignment.}
    \label{fig:obs_model_Sep}
\end{figure*}

\subsubsection{Interstellar radiation field}
The spectral energy density of the ISRF comprises the stellar radiation and cosmic microwave background (CMB), which can be described by (see HD23 and references therein)
\begin{equation} \label{eq:urad}
    u_\lambda = U_{\rm scale} \times \left[u^{\rm UV}_{\lambda} + \sum^{3}_{i=1} \frac{4\pi}{c}W_{i}B_{\lambda}(T_{i})\right] + \frac{4\pi}{c}B_{\lambda}(2.7\,\rm K),
\end{equation}
where $u^{\rm UV}_{\lambda}$ is the UV component of the field (see Equation 10 in HD23), in which $u^{\rm UV}_{\lambda}(\lambda < 912\,\AA)=0$ (the EUV radiation field is excluded). The stellar temperatures and the dilute factors are $T_{1}=7500\,$K, $T_{2}=4000\,$K, $T_{3}=3000\,$K, $W_{1}=10^{-14}$, $W_{2}=1.65\times 10^{-13}$ and $W_{3}=7\times 10^{-13}$. 

The dimensionless radiation strength and mean wavelength of the radiation spectrum are calculated as
\begin{equation}\label{eq:lambda_mean}
    \begin{split}
        U&=\int^{20\,\rm \mu m}_{0.091\,\rm \mu m}u_{\lambda}d\lambda/u_{\rm MMP83}, \\
        \bar{\lambda} &= \frac{\int^{20\,\rm \mu m}_{0.091\,\rm \mu m}\lambda u_\lambda d\lambda}{\int^{20\,\rm \mu m}_{0.091\,\rm \mu m} u_\lambda d\lambda} \simeq 1.2\,\rm \mu m, \\
    \end{split}
\end{equation}
with $u_{\rm MMP83}=8.64\times 10^{-13}\,\rm erg\,cm^{-3}$ (\citealt{1983A&A...128..212M}). For a typical ISRF, $U=1$. The dust temperature at equilibrium is estimated from the radiation as $T_{\rm dust}=16.4U^{1/6}(a/10^{-5}\,\rm cm)^{-1/15}\,\rm K$. We note that $\bar{\lambda}$ is independent of $U_{\rm scale}$ due to cancellation of $U_{\rm scale}$ in Equation (\ref{eq:lambda_mean}).

\subsubsection{Effect of only DG alignment on the super-Serkowski spectra} \label{sec:only_DG}
To investigate the impact of the alignment of DG, we fix a set of parameters of $U_{\rm scale}=1$ (i.e. $\lambda=1.2\,\mu$m), $n_{\rm H}=30\,\rm cm^{-3}$, $T_{\rm gas}=100\,$K, and $\gamma=0.1$ (this set of parameters mimics the CNM condition), which results in $a_{\rm align}=0.05\,\mu$m. We then calculate the value of $f_{\rm min}(a)$ from Equation \ref{eq:fmin} embedded with another free parameter, $\alpha_{\rm DG}$. 

In total, we have seven free parameters, including $f_{\rm max}$, $\alpha_{\rm DG}$, three parameters for the grain size distribution of \textsc{astrodust} ($A_{1}$, $A_{2}$ and $A_{3}$) and two free parameters ($B_{1}$ and $B_{2}$) for the PAH size distribution. We adopted the values of the gas column density of $1.23\times 10^{21}\,\rm cm^{-2}$ for HD 30614 based on \cite{2021ApJS..257...63Z}, while varying them as an additional free parameter for HD 204827.

Figure \ref{fig:obs_model_supSep}(a) shows the best fits of our model (the shaded area corresponds to the parameters with the top 500 posterior distributions with the best posterior showing by the solid magenta line) to the observations for those with the super-Serkowski spectrum in HD 30614 and HD 204827. Overall, our model successfully reproduces the observed extinction curve, including the 2175$\AA$ bump; however, the predicted polarisation spectra differ from the observations for $\lambda^{-1}>3\,\rm \mu m^{-1}$, with the model underestimating the degree of polarisation.

\subsubsection{Effect of only enhanced RATs on the super-Serkowski spectra} \label{sec:onlyRAT_CNM}
To investigate the role of the effectiveness of RAT alignment, we excluded DG alignment by fixing $f_{\rm min}=0$ and varying $a_{\rm align}$ by setting $U_{\rm scale}$ as a free parameter (other free parameters are the same as in Section \ref{sec:only_DG}). In this case, we have seven free parameters: $f_{\rm max}$, $U_{\rm scale}$, $A_{1}$, $A_{2}$, $A_{3}$, $B_{1}$ and $B_{2}$.

Figure \ref{fig:obs_model_supSep}(b) shows the best models for $f_{\rm min}=0$. The extinction curve is nicely reproduced and is better responsible for the slope in the mid-UV in the polarisation spectrum. However, the best model tends to underestimate the degree of polarisation towards the highest wavenumber. Our best model corresponds to the values of $a_{\rm align}\simeq 0.017\,\rm \mu m$ for HD 30614 and $a_{\rm align}\simeq 0.032\,\rm \mu m$ for HD 204827, which are smaller than $0.05\,\mu$m, indicating that the RATs must be enhanced to align smaller grains. To achieve this best fit, the best radiation intensity must be $U\sim 10^{4}$ for HD 30614 and $U\sim 85$ for HD 204827 times the typical ISRF, which could be difficult to reconcile in the CNM term with $n_{\rm H}=30\,\rm cm^{-3}$ and $T_{\rm gas}=100\,$K.

\subsubsection{Effect of a combination of paramagnetic and enhanced RAT alignments on the super-Serkowski spectrum}
We combine both the effects of the RAT and DG alignments in Figure \ref{fig:obs_model_supSep}(c) allowing the variation of $U_{\rm scale}$ and $\alpha_{\rm DG}$. Therefore, the joint effect of $a_{\rm align}$ and $f_{\rm min}$ can be incorporated. In this case, we have eight free parameters, and $f_{\rm min}$ and $a_{\rm align}$ are consistently determined. 

For HD 30614, the best models for both both extinction curve and polarisation spectrum are quite similar to these of panel (b). For HD 204827, the shape of the best model similar to panel (b), with a lower deviation for $\lambda^{-1}>5\,\rm \mu m^{-1}$. However, similar to Section \ref{sec:onlyRAT_CNM}, the best radiation intensities are extremely high, $U \sim 10^{4}$ for HD 30614 and even higher with $U\sim 10^{3}$ for HD 204827, which could not be satisfied within the CNM condition.

\subsection{Interpretation of super-Serkowski spectra with the Per OB3 super-bubble} \label{sec:SuSep_OB3}
One special feature is that some super-Serkowski spectra among WUPPE targets pass through the Per OB3 super-bubble, another name is $\alpha$ Persei (see e.g. \citealt{1995ApJ...445..947C} and \citealt{2022Ap&SS.367..127A}). As a result, the second hypothesis for the origin of super-Serkowski spectra is that the extreme ultraviolet radiation with $\lambda<912\,\AA$ and stronger radiation fields from this super-bubble can help smaller grains to align by RATs, and thus only the RAT alignment can account for the observed super-Serkowski spectrum. To test this hypothesis, we re-perform the fit to the polarisation spectrum and extinction curve using the RAT alignment with ($f_{\rm min}=DG$) and without ($f_{\rm min}=0$) the DG alignment.

\subsubsection{Physical parameters for the Per OB3 super-bubble}
Based on \cite{2012ApJ...752...58Z}, of the 183 stars in their catalogue of Per OB3 members, 29 are B stars, with the two earliest being B3 V. Hence, the B5 star is a reasonable first-order template. At a distance $r$, the spectral energy density of the radiation from the radiation source in this super-bubble approximates as
\begin{equation}
    u_{\lambda} = U_{\rm scale} \times \frac{(4\pi R^{2}_{\ast}) \pi B_{\lambda}(T_{\ast})}{4\pi r^{2}c}, 
\end{equation}
where $R_{\ast}$ is the stellar radius and $U_{\rm scale}$ is the scaled parameter. In this case, we can define that the dimensionless radiation strength, which accounts for the EUV, is $U=\int_{1\AA}^{\infty}u_{\lambda}d\lambda/u_{\rm MMP83}$. For $T_{\ast}\simeq 15000\,$K, the mean wavelength of the radiation, independent of the distance $r$, is
\begin{equation}
    \bar{\lambda}=\frac{\int_{1\AA}^{\infty}\lambda u_{\lambda}d\lambda}{\int_{1\AA}^{\infty} u_{\lambda}d\lambda} \simeq 0.36\,\rm \mu m.
\end{equation} 
The B stars are scattered within the Per OB3, and thus the distance $r$ from each of these stars to our target remains unknown. For simplicity, we adopt a representative scenario in which the radiation source is in the centre of Per OB3 with a radius of $30\,$ pc and $R_{\ast}=30\,R_{\odot}$. The contributions of other types of stars are neglected. 

The gas density within the bubble is relatively low, in which we use $n_{\rm H}=2\,\rm cm^{-3}$ for HD 30614 and $n_{\rm H}=30\,\rm cm^{-3}$ and the gas temperature is relatively high $T_{\rm gas}=10000\,$K, according to \cite{1977ApJ...218..377W}. The reason for different values of the gas density is discussed in Section \ref{sec:ext_dens}. The main reason is that the best path length is unrealistic when adopting $n_{\rm H}=2\,\rm cm^{-3}$. We used the magnetic field measurements using Faraday rotation (and dispersion measures) from \citet{sobey2019} combined with the pulsar distances of \citet{ronchi2021} to estimate the magnetic field strength in the Per OB3 super-bubble. The details are described in Section \ref{sec:Bfield_perOB3}. The strength of the magnetic field approximates $5\,\mu$G. 

\subsubsection{Results}
Figure \ref{fig:HD30614_EUV} illustrates the best model fits for HD 30614 (\textit{panel a}) and HD 204827 (\textit{panel b}). It is clear that the fits obtained with and without DG alignment are nearly identical for $\lambda^{-1}<5.5\,\rm \mu m^{-1}$, but begin to diverge noticeably beyond this value, i.e. the spectral slope is positive when the DG alignment is taken into account; otherwise it is negative, and the model does not converge well in this far-UV region because no data are available there. These indicate that within the wavelength coverage used in this work, the super-Serkowski spectrum can primarily be attributed to the RAT alignment, which is due to the influence of the EUV or more intense radiation fields. However, a distribution of the DG alignment towards the longest wavenumber improved the fit quality, especially the last data point at $\lambda^{-1}\simeq 6\,\rm \mu m^{-1}$ and hence the effect of the DG alignment becomes more important at extended wavelengths.

We note that the above discussion assumed that the total dust along the LOS is concentrated in the super-bubble. In realistic situations, there may exist several dust layers along the LOS, and the exact contribution of dust in the super-bubble to the super-Serkowski spectra remains uncertain.

\subsection{Interpretation of Serkowski spectra}
\subsubsection{Observational targets}
The sight lines exhibiting the Serkowski spectra include HD 37903 and HD 161056. The observed polarisation spectra were adopted from \cite{1995ApJ...445..947C}. The extinction curve and the gas column density of $N_{\rm H}=3.16\times 10^{21}$ for HD 37903 were adopted from \cite{2021ApJS..257...63Z}, which were derived similarly to that of HD 30614. For HD 161056, we were unable to find the extinction curves from the literature; therefore, we fit solely to the polarisation spectrum. Therefore, we fix the parameters $B_{1}=7.52\times 10^{-7}$ and $B_{2}=8.09\times 10^{-10}$ for PAH as in HD23, and vary the free parameter $N_{\rm H}$. 

For these targets, we used the radiation field scaled from ISRF as in Equation \ref{eq:urad} and the physical conditions of CNM as in Section \ref{sec:supSep_CNM} in which $n_{\rm H}=30\,\rm cm^{-3}$, $T_{\rm gas}=100\,$ K, $\bar{\lambda}=1.2\,\mu$m, and $\gamma=0.1$. Depending on whether the DG alignment is taken into account, the number of free parameters varies. For HD 37903, there are eight and seven free parameters respectively with and without the DG alignment, which is seven or six for HD 161056.

\subsubsection{Results}
Figure \ref{fig:obs_model_Sep} shows the comparison between our best models with the effect of thean DG alignment incorporated with the RAT alignment at a fixed $a_{\rm align}=0.05\,\mu$m (panel a) and with only the RAT alignment (panel b). Interestingly, these two distinct cases identically reproduce the observed spectra. When the DG alignment acts as a play, the best IR damping factor $\alpha_{\rm DG}$ is required to be negligible with $\alpha_{\rm DG} \sim 10^{-2}$, showing that the DG alignment is very inefficient. Therefore, only the RAT alignment sufficiently accounts for the Serkowski spectra. The contribution of DG alignment is not obvious.

\section{Discussion} \label{sec:discussion}
\subsection{Whether only RAT alignment can account for the polarisation at UV bands}
In this work, we considered a small sample of sightlines exhibiting both super-Serkowski and Serkowski spectra. The RAT alignment nicely reproduces the polarisation spectrum from near-IR to UV with $\lambda^{-1} < 3 \,\rm \mu m^{-1}$. The polarisation at a longer wavenumber is the subject of discussion.

For the super-Serkowski spectrum, Figures \ref{fig:obs_model_supSep} and \ref{fig:HD30614_EUV} indicate that the RAT alignment for a typical interstellar medium with $a_{\rm align}=0.05\,\mu$m, the variation of the DG alignment could not explain the excess of the degree of polarisation at mid-UV and the spectral shape of the observations of our samples. When the DG alignment is excluded and only the enhancement of the RAT alignment is considered, the slope is better fitted, but with a lower degree of polarisation predicted towards the highest wavenumber ($\lambda^{-1} \geq 5.5\,\rm \mu m^{-1}$). When both the enhanced RAT and the DG alignments are integrated, the fit at these highest wavenumbers is better, showing that the influence of DG alignment becomes more pronounced for such long wavenumbers.

The enhancement of the RAT alignment is considered under two physical conditions. The first scenario mimics the CNM with $n_{\rm gas}=30\,\rm cm^{-3}$, $T_{\rm gas}=100\,$K, $\bar{\lambda}=1.2\,\mu$m, and $\gamma=0.1$. The best model with RAT requires $U\sim 10^{4}$ for HD 30614 and $U\sim 10^{2}$ for HD 204827, which to us seems very difficult to reconcile with the CNM condition. The second scenario mimics the interstellar super-bubble with $T_{\rm gas}=10000\,$K, $\bar{\lambda}=0.36\,\mu$m and $\gamma=1$. The best radiation is also high, on the order of $10^{2}$, but this extremely intense radiation intensity might be aligned with the EUV radiation field of the bubble. Therefore, the super-Serkowski spectrum associated with the super-bubble is our preferred scenario.

For the super-bubble scenario, if the distance $r$ from the irradiation source is shorter than 30$\,$pc that we used, the radiation intensity is more intense, causing more grains to align (the value of $a_{\rm align}$ is smaller). Consequently, using the parameters as in Table \ref{tab:params_derived}, the excess in the UV bands becomes enhanced, over the observed values. However, the best fits are also achieved for a less efficient DG alignment (lower $a_{\rm DG}$). However, the DG alignment is still important towards the highest wavenumbers.

For the Serkowski spectrum, Figure \ref{fig:obs_model_Sep} explicitly demonstrates that only RAT alignment by itself within typical CNM conditions can nicely explain the observed spectrum, and the contribution of DG alignment is minimal.

\subsection{Whether the RAT alignment degree changes from Serkowski to super-Serkowski spectra} \label{sec:special_condition}
Within the RAT paradigm, the alignment function (Eq.\ref{eq:falign}) is completely described by two key parameters, including the alignment size, $a_{\rm align}$, and the maximum degree of alignment of large grains, $f_{\rm max}$ (see \citealt{2022FrASS...9.3927T}). Both parameters are theoretically predicted to vary with local environments, including gas properties and radiation field \citep{HoangLaz.2008,2021ApJ...908...12L,2025ApJ...994..115H}.

The best-fit parameters obtained from fitting the Serkowski or super-Serkowski curves can provide insight into this theoretical prediction. Table \ref{tab:params_derived} shows two important features. First, the values of $U$ are significantly higher for the LOS with a super-Serkowski curve than for the LOS with a Serkowski curve, corresponding to a smaller $a_{\rm align}$ and a broader range of aligned grain sizes for a given $a_{\rm max}$. Taking two cases of HD 30614 and HD 37903, the values of $f_{\rm max}$ are significantly decreased from $0.84$ to $0.5$ in the super-Serkowski sight lines. This may be evidence of a reduction in the degree of alignment in strong radiation, as found in \cite{2025ApJ...994..115H}. Specifically, it is shown that, in the radiation-dominated region where $U\left(\frac{n_{\rm H}}{10\,\rm cm^{-2}}\right)^{-1}\left(\frac{T_{\rm gas}}{100\,\rm K}\right)^{-1}>1$, $f_{\rm max}$ can be reduced by a factor of 2-3 due to the trap of the majority of grains rotating at low angular momentum (RAT trapping), while in the collision-dominated region where $U\left(\frac{n_{\rm H}}{10\,\rm cm^{-2}}\right)^{-1}\left(\frac{T_{\rm gas}}{100\,\rm K}\right)^{-1}<1$, $f_{\rm max}$ could even reach 1 because collisional and magnetic excitations can slowly transport grains from the state of low angular momentum to high angular momentum. However, the magnetic field geometry along these LOS is not the same, which prevents us from a strong conclusion on the reduction of $f_{\rm max}$. However, it suggests the need to accurately constrain $f_{\rm max}$ using physics-based modelling for a given physical condition.

\subsection{Extinctions and gas density} \label{sec:ext_dens}
The best models to the extinction curve towards the LOSs with the super-Serkowski spectra, shown in Figures \ref{fig:obs_model_supSep} or \ref{fig:HD30614_EUV}, exhibit an underestimate of the extinction for $\lambda^{-1}> 5\,\rm \mu m^{-1}$ (right after the 2175$\AA$ peak). We conducted a series of tests and found that this underestimation is not due to physical parameters. A similar mismatch when using \textsc{astrodust} could also be seen in Figure 4 in \cite{2023ApJ...948...55H}. Furthermore, the fitting is much better for the LOSs with Serkowski spectra (see Figure \ref{fig:obs_model_Sep}). Therefore, we suspect that this issue is probably related to the grain composition and this composition is different between the Serkowski and super-Serkowski sightlines. 

In Section \ref{sec:SuSep_OB3}, we adopted two different values for the gas volume density: $n_{\rm H}=2\,\rm cm^{-3}$ for HD 30614. With this value, a simple path length estimate, $L=N_{\rm H}/n_{\rm H}$, yields roughly 195$\,$pc, which is a plausible size for a super-bubble (see e.g. \citealt{2023MNRAS.523.5995M}). If we apply the same volume density to HD 204827, we can still obtain a statistically acceptable fit (see Figure \ref{fig:supSep_bubble_unreal}b), but the inferred path length is $L \simeq 2\times 10^{22}\,\rm cm^{-2}/2\,\rm cm^{-3} \simeq 3.2\,\rm kpc$, which is likely unphysical (the distance to HD 204827 is less than 1$\,$kpc according to the Gaia parallax measurement of 1.077$\,$mas). We therefore assume a plausible denser medium, $n_{\rm H}=30\,\rm cm^{-3}$, for HD 204827, supported by \cite{2023MNRAS.523.5995M}, which reduces the path length to a more realistic value of about 200$\,$pc. When we apply this higher volume density to HD 30614, the fit quality does not change significantly (see Figure \ref{fig:supSep_bubble_unreal}); however, the degree of polarisation at the highest wave number, which is beyond the observed range, deviates less from the serkowski curve, which arises purely from the more efficient rotational damping of the DG alignment at higher densities.

Furthermore, regardless of the inuniformity of $n_{\rm H}$ along the LOS, according to Equation \ref{eq:pext}, there are three possibilities to increase the value of $p_{\rm ext}$, one of which is due to the intrinsic properties of the dust grains: the elongation. Thus, we perform a test by increasing the grain elongation with the aspect ratio of 1.4 to an extreme case of 3, where the value of $N_{\rm H}$ is reduced only by a factor of 3.3. However, a highly elongated grain within the diffuse ISM is unlikely to be physical. Therefore, the shape of the grain does not provide a direct explanation, which left the hypothesis of having different amounts of gas and dust in different super-Serkowski sightlines. Within this simple test, one probably indicates that different super-Serkowski curves probe different parts of the OB3 bubble.  

\begin{figure}
    \includegraphics[width=1.0\linewidth]{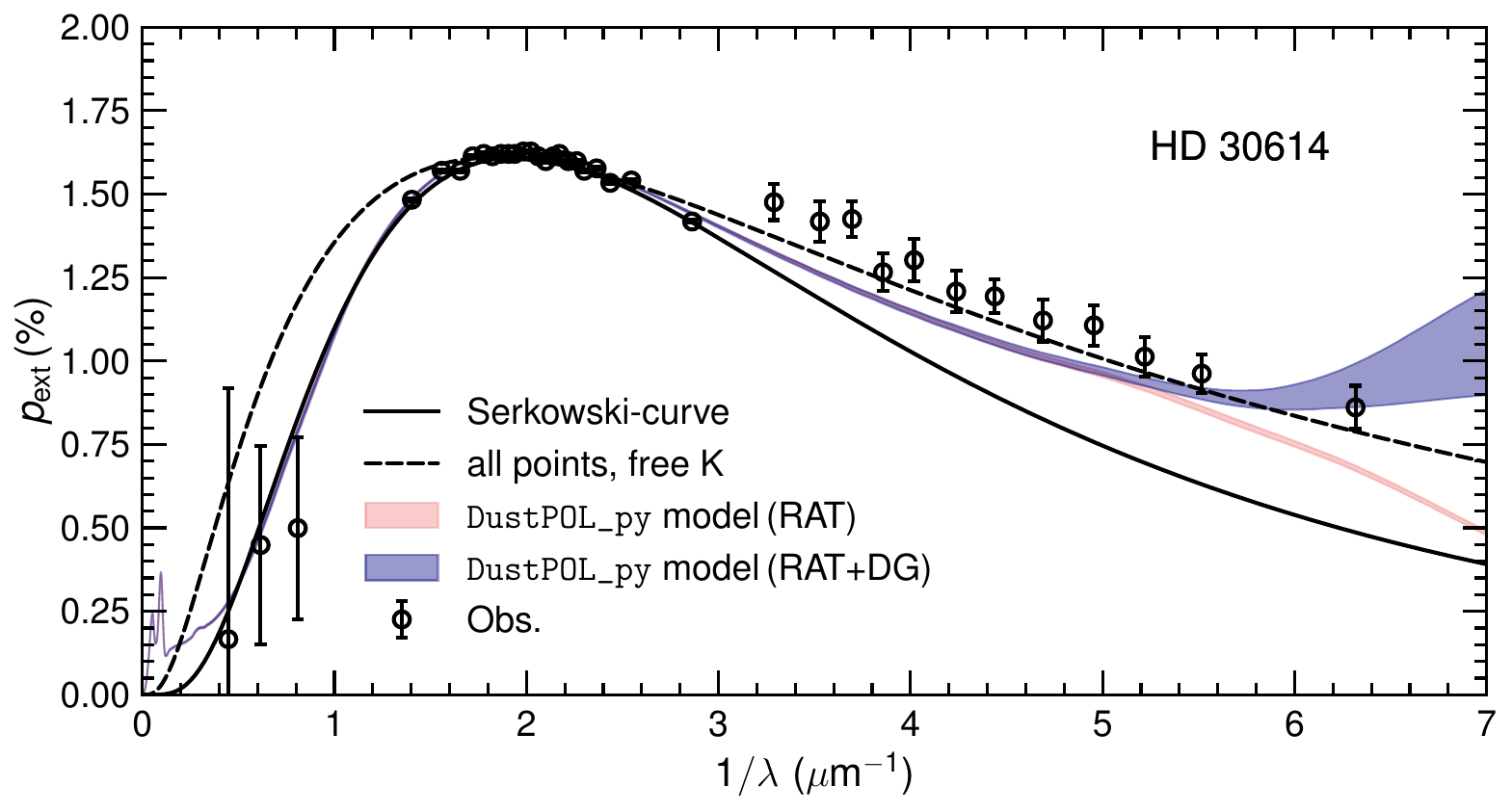}
    \caption{Polarisation spectrum towards HD 30614, similar to Figure \ref{fig:HD30614_EUV}. We show the Serkowski curve fitted to all data points (black dashed line) for a comparison. Infrared polarimetry is crucial in assessing whether the spectrum follows the Serkowski curve or deviates into a super-Serkowski curve.}
    \label{fig:Sers.vs.supSers}
\end{figure}
\subsection{Importance of spectro-polarimetry from near-IR to mid-UV}
As shown in Figures \ref{fig:obs_model_supSep} and \ref{fig:obs_model_Sep}, multi-wavelength observations in the UV bands for $\lambda^{-1}>2\,\rm \mu m^{-1}$ are essential to distinguish whether the Serkowski curve is followed or an access appears. Each of these different spectra reflects the distinct physical conditions and physics of grain alignment. Furthermore, the super-Serkowski spectra exhibited in 9/28 sightlines observed by WUPPE and HST/FOS. Therefore, future emissions with spectro-polarimetry capacity at UV bands are required to consolidate our understanding of the alignment of dust grains and ultimately advance the constraints on their physical properties. Several missions have been proposed, for example, LUVOIR/POLLUX (\citealt{2018AAS...23141901B}), Polstar (\citealt{2022Ap&SS.367..127A}), and PUFFINS (\citealt{10.1117/12.3065702}). 

Based on the findings reported in \cite{1995ApJ...445..947C}, applying the Serkowski formulae to fit the super-Serkowski spectra of starlight polarisation results in an excess at near-infrared (near-IR) wavelengths. Consequently, without the constraints provided by near-IR polarimetry, there is a risk of bias in determining whether the Serkowski or super-Serkowski spectra are more appropriate. The underlying reason is that the mid-UV wavelength excess in the super-Serkowski spectra broadens the profile, leading to a broader best-Serkowski curve in order to match the data accurately. Although this best Serkowski curve may account for the degree of polarisation at UV wavelengths, it undoubtedly leads to an overestimation in the near-IR range. This need for near-IR polarimetries is shown in Figure \ref{fig:Sers.vs.supSers}, which illustrates the comparison between our best model and the Serkowski fit to all data points by the black dashed line. Similar observations can be seen in other starlights with super-Serkowski spectra. Furthermore, this comparison strengthens the ability of our model in explaining the full spectrum of starlight polarisation.

\begin{figure*}
    \centering
    \includegraphics[width=1.0\linewidth]{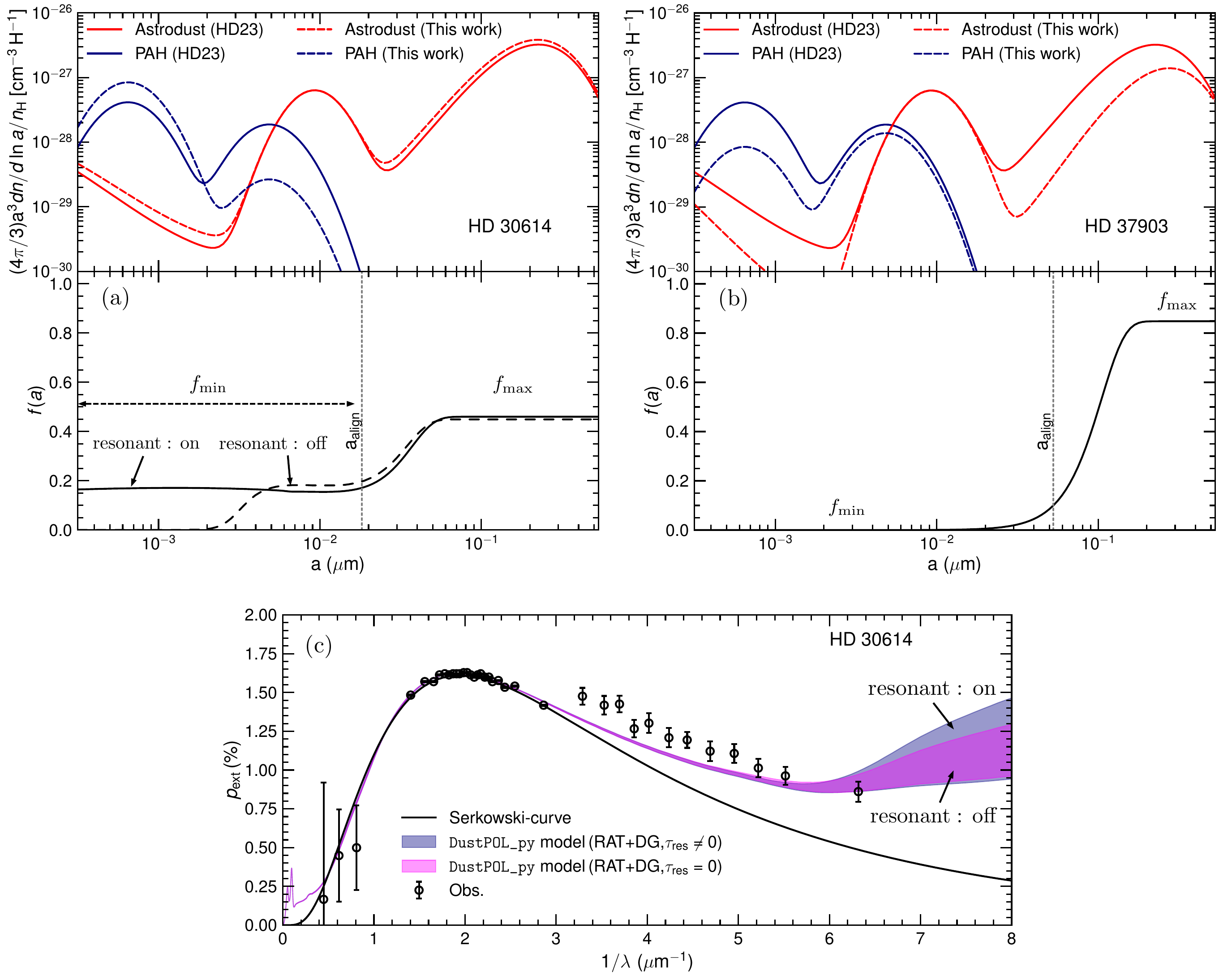}
    \caption{{\it Panel a-b}: Distribution of grain mass (top) and the alignment function (bottom) for the HD 30614 (within Per OB3 bubble) and HD 37903. With respect to HD 39703, the amount of VSGs ($a\leq 10^{-3}\,\rm \mu m$) is more abundant, while the value of $f_{\rm max}$ is lower in HD 30614. The alignment function for $a<a_{\rm align}$ of HD 30614 is also more complex, in which the resonant alignment for VSGs and paramagnetic alignment for larger grains are efficient. {\it panel c}: Super-Serkowski spectrum in HD 30614 and the best model with and without the resonant alignment for VSGs. The enhanced alignment for VSGs results in a higher degree of polarisation but for $\lambda^{-1}\geq 6\,\rm \mu m^{-1}$, and the absence of data beyond this range results in poor convergence of the model.}
    \label{fig:dnda}
\end{figure*}
\subsection{Importance of dust polarisation at far-UV}
From the best models in Figure \ref{fig:obs_model_supSep} (for the super-Serkowski spectrum with the super-bubble condition) and Figure \ref{fig:obs_model_Sep} (for the Serkowski spectrum), we show the dust mass distribution and the alignment function for two representative cases of HD 30614 and HD 37903, which are shown in Figure \ref{fig:dnda}(a,b). The comparison is visualised through the HD23 distribution. The mass distribution of the aligned \textsc{astrodust} in HD 30614 is more abundant than HD 37903 for both large grains ($a>a_{\rm align}$) and VSGs ($a<10^{-3}\,\rm \mu m$). The alignment function is also distinct. For HD 37903, $f(a)\rightarrow 0$ when $a\ll a_{\rm align}$ because DG alignment is not effective, whereas the situation is more complicated for HD 30614. In this case, paramagnetic alignment significantly improves the alignment of grains with $2\times 10^{-3}\,\rm \mu m <a<a_{\rm align}$ with $\max(f(a))\simeq 0.2$, whereas $f(a)=0$. Resonant alignment allows for effective alignment of VSGs, achieving $f(a)\simeq 0.2$ from the fitting, and without this alignment process, $f(a)$ disappears for these VSGs, as shown in Figure \ref{fig:R_DG}.

Figure \ref{fig:dnda}c illustrates the comparison of the HD 30614 spectrum with and without resonant alignment. When this alignment is active, $p_{\rm ext}$ is larger than when it is inactive, at the very shortest wavelengths ($\lambda^{-1}\geq 6.5\,\rm \mu m^{-1}$), although the best-fit model does not converge well in this wavelength range due to the lack of observed data. Within the wavelength coverage used in this work, the contribution of resonant alignment is minimal. However, we note that we fixed $a_{\rm min}\simeq 3\times 10^{-4}\,\rm \mu m$. In reality, this value might be higher because the smallest grains might not survive the destruction processes, for example the thermal sublimation, sputtering, or rotational destruction by the supersonic drag between the charged grains and neutral gas. All of these can be possible within the physical conditions of the super-bubble, like Per OB3. Therefore, observations with much shorter wavelengths in the far-UV spectrum can be an essential tool to study the alignment physics and evolution of VSGs.

\newpage
\section{Summary} \label{sec:summary}
In this study, we model the extinction curve and polarisation spectrum of starlight ranging from infrared to far-ultraviolet based on the RAT alignment and DG alignment theories. For grains with $a>a_{\rm align}$ (minimum size aligned by RAT), the RAT alignment is dominant and the internal alignment is assumed to be perfect. The degree of net alignment is parameterised by $f_{\rm max}$. For grains where $a<a_{\rm align}$, the alignment of the DG is predominant and the degree of alignment is derived from a simple but physically based model. The determination of $a_{\rm align}$ is considered in three scenarios: a typical ISRF with the Lyman limit, an enhanced ISRF with the Lyman limit, and a super-bubble radiation field with the EUV component.

We apply our model to simultaneously interpret the observations for five stars, such as HD 37903, HD 161056, HD 30614, and HD 204827. The first two targets show Serkowski spectra, while the last three show the super-Serkowski feature. The Interpretations simultaneously integrate the extinction curves observed for three targets (except for HD 161056).

\begin{itemize}
    \item[1-] For the super-Serkowski spectra, we consider two scenarios.
    \begin{itemize}
        \item In the first scenario, we adopt the physical conditions of the CNM with the radiation field following the standard ISRF's shape cutoff below 912$\,\AA$ and its intensity scaled by the ISRF. Our model nicely reproduces the excess phenomenon in the observed polarisation spectra and highlights that the RAT alignment itself or in combination with the DG reasonably accounts for both observations for $\lambda^{-1} < 5.5 \,\rm\mu m^{-1}$. The best radiation intensity is extremely high, scaled to the ISRF (on the order of $10^{4}$ for HD 30614 and $10^{2}$ for HD 204827).

        \item In the second scenario, we follow the hypothesis that starlight revealed that the super-Serkowski spectrum seen in the WUPPE observations passes through the Per O3 super-bubble, for which the EUV radiation field contributes to the alignment of grains. We then use the physical parameter for an interstellar super-bubble. Similarly to the previous scenario, the RAT alignment alone can nicely account for the super-Serkowski spectra, and the distribution of the DG alignment becomes pronounced for $\lambda^{-1}\geq 5.5\,\rm \mu m^{-1}$. The maximum values of the radiation field reach about $10^{2}$ for HD 30614 and several tens for HD 204827.
        
        \item When comparing these two scenarios, the predicted slope of the polarisation spectrum from our model in the second scenario aligns more closely with the observed data. Additionally, the combination with high radiation intensity with the intense radiation features of $\bar{\lambda}$ and $\gamma$ in the super-bubble appears more physically plausible.

        \item Although our model can account for the excess in the polarisation spectrum for ($3<\lambda^{-1}<5\,\rm \mu m^{-1}$), our best model always underestimates the actual observed polarisation fraction in this range. Investigating the physical underneath is required.
    \end{itemize}
       
    \item[2-] For the Serkowski spectra, the RAT alignment under a typical CNM conditions nicely reproduces the observations with an insignificant contribution by the DG alignment.

    \item[3-] Furthermore, it is observed that the radiation intensity values $U$, and the maximum alignment efficiency of RAT $f_{\rm max}$ might show an inverse relationship, where a higher $U$ is associated with a reduced $f_{\rm max}$ (HD 30614 versus HD 37903). This phenomenon might align with the impact of radiative trapping on grain rotation and requires further research.

    \item[4-] Multiple wavelengths from near-IR to far-UV are essentially needed to confirm whether the polarisation falls into the Serkowski or super-Serkowski spectrum, and it also provides a promising tool for studying the physical properties and evolution of nanosized dust.
\end{itemize}

To enhance our understanding of grain alignment physics, polarimetric observations in the UV spectrum that gather observations of larger samples are particularly needed. Our work can contribute and benefit from some of the potential future missions, such as the small-sat PUFFINS (\citealt{10.1117/12.3065702}) or the HWO/POLLUX spacecraft (\citealt{2018AAS...23141901B}).

\begin{acknowledgement}
T.H. acknowledge the support from the main research project (No. 2025186902) from Korea Astronomy and Space Science Institute (KASI). This work was partially supported by a grant from the Simons Foundation to IFIRSE, ICISE (916424, NH). The authors thank ICISE staff for excellent support and hospitality. L.N.T would like to express his gratitude to Prof. Karl M. Menten and Dr. Yannick Giraud-Heraud, who have sadly passed away, for their mentor, support and encouragement.
\end{acknowledgement}

\bibliographystyle{aa} 
\bibliography{bib}

\begin{thebibliography}{74}
\expandafter\ifx\csname natexlab\endcsname\relax\def\natexlab#1{#1}\fi

\bibitem[{Anche {et~al.}(2025)Anche, Kang, Gorkom, Vargas, Chung, Spitzer,
  Kupinski, Andersson, Clayton, Douglas, Fossati, Gasho, Gopinathan, Hamden,
  Hoang, Klupar, Lau, Lazarian, Le, Rosenbluth, Suresh, \&
  Vargas}]{10.1117/12.3065702}
Anche, R.~M., Kang, H., Gorkom, K.~V., {et~al.} 2025, in Polarization Science
  and Remote Sensing XII, ed. M.~K. Kupinski \& J.~A. Shaw, Vol. 13615,
  International Society for Optics and Photonics (SPIE), 136150K

\bibitem[{{Andersson} {et~al.}(2022){Andersson}, {Clayton}, {Doney},
  {Panopoulou}, {Hoang}, {Magalhaes}, {Yan}, {Ignace}, \&
  {Scowen}}]{2022Ap&SS.367..127A}
{Andersson}, B.~G., {Clayton}, G.~C., {Doney}, K.~D., {et~al.} 2022, \apss,
  367, 127

\bibitem[{{Andersson} {et~al.}(2015){Andersson}, {Lazarian}, \&
  {Vaillancourt}}]{2015ARA&A..53..501A}
{Andersson}, B.~G., {Lazarian}, A., \& {Vaillancourt}, J.~E. 2015, ARAA, 53,
  501

\bibitem[{{Andersson} {et~al.}(2013){Andersson}, {Piirola}, {De Buizer},
  {Clemens}, {Uomoto}, {Charcos-Llorens}, {Geballe}, {Lazarian}, {Hoang}, \&
  {Vornanen}}]{bga2013}
{Andersson}, B.-G., {Piirola}, V., {De Buizer}, J., {et~al.} 2013, \apj, 775,
  84

\bibitem[{{Andersson} {et~al.}(2011){Andersson}, {Pintado}, {Potter},
  {Strai{\v{z}}ys}, \& {Charcos-Llorens}}]{2011A&A...534A..19A}
{Andersson}, B.~G., {Pintado}, O., {Potter}, S.~B., {Strai{\v{z}}ys}, V., \&
  {Charcos-Llorens}, M. 2011, \aap, 534, A19

\bibitem[{{Bhatt}(2000)}]{bhatt2000}
{Bhatt}, H.~C. 2000, \aap, 362, 715

\bibitem[{{Bouret} {et~al.}(2018){Bouret}, {Neiner}, {Lopez Ariste},
  {Viv{\`e}s}, {Muslimov}, {Lopes}, {Costeraste}, {Brachet}, \& {POLLUX
  Consortium}}]{2018AAS...23141901B}
{Bouret}, J.-C., {Neiner}, C., {Lopez Ariste}, A., {et~al.} 2018, in American
  Astronomical Society Meeting Abstracts, Vol. 231, American Astronomical
  Society Meeting Abstracts \#231, 419.01

\bibitem[{{Cardelli} {et~al.}(1989){Cardelli}, {Clayton}, \&
  {Mathis}}]{1989ApJ...345..245C}
{Cardelli}, J.~A., {Clayton}, G.~C., \& {Mathis}, J.~S. 1989, \apj, 345, 245

\bibitem[{{Chiar} {et~al.}(2006){Chiar}, {Adamson}, {Whittet}, {Chrysostomou},
  {Hough}, {Kerr}, {Mason}, {Roche}, \& {Wright}}]{chiar2006}
{Chiar}, J.~E., {Adamson}, A.~J., {Whittet}, D.~C.~B., {et~al.} 2006, \apj,
  651, 268

\bibitem[{{Clayton} {et~al.}(1995){Clayton}, {Wolff}, {Allen}, \&
  {Lupie}}]{1995ApJ...445..947C}
{Clayton}, G.~C., {Wolff}, M.~J., {Allen}, R.~G., \& {Lupie}, O.~L. 1995, \apj,
  445, 947

\bibitem[{{Clayton} {et~al.}(2003){Clayton}, {Wolff}, {Sofia}, {Gordon}, \&
  {Misselt}}]{clayton2003}
{Clayton}, G.~C., {Wolff}, M.~J., {Sofia}, U.~J., {Gordon}, K.~D., \&
  {Misselt}, K.~A. 2003, \apj, 588, 871

\bibitem[{{Codina-Landaberry} \& {Magalhaes}(1976)}]{codina1976}
{Codina-Landaberry}, S. \& {Magalhaes}, A.~M. 1976, \aap, 49, 407

\bibitem[{{Coyne} \& {Gehrels}(1967)}]{coyne1967}
{Coyne}, G.~V. \& {Gehrels}, T. 1967, \aj, 72, 887

\bibitem[{{Crutcher} {et~al.}(2003){Crutcher}, {Heiles}, \&
  {Troland}}]{crutcher2003}
{Crutcher}, R., {Heiles}, C., \& {Troland}, T. 2003, in Turbulence and Magnetic
  Fields in Astrophysics, ed. E.~{Falgarone} \& T.~{Passot}, Vol. 614, 155--181

\bibitem[{{Crutcher} {et~al.}(2010){Crutcher}, {Wandelt}, {Heiles},
  {Falgarone}, \& {Troland}}]{2010ApJ...725..466C}
{Crutcher}, R.~M., {Wandelt}, B., {Heiles}, C., {Falgarone}, E., \& {Troland},
  T.~H. 2010, \apj, 725, 466

\bibitem[{{Davis} \& {Greenstein}(1951)}]{DG1951}
{Davis}, Leverett, J. \& {Greenstein}, J.~L. 1951, \apj, 114, 206

\bibitem[{{Dolginov} \& {Mitrofanov}(1976)}]{1976Ap&SS..43..291D}
{Dolginov}, A.~Z. \& {Mitrofanov}, I.~G. 1976, Ap\&SS, 43, 291

\bibitem[{{Draine} \& {Fraisse}(2009)}]{2009ApJ...696....1D}
{Draine}, B.~T. \& {Fraisse}, A.~A. 2009, ApJ, 696, 1

\bibitem[{Draine \& Lazarian(1998)}]{draine_1998}
Draine, B.~T. \& Lazarian, A. 1998, ApJ, 508, 157

\bibitem[{{Draine} \& {Lazarian}(1998)}]{1998ApJ...508..157D}
{Draine}, B.~T. \& {Lazarian}, A. 1998, \apj, 508, 157

\bibitem[{{Draine} \& {Li}(2007)}]{2007ApJ...657..810D}
{Draine}, B.~T. \& {Li}, A. 2007, \apj, 657, 810

\bibitem[{{Draine} \& {Weingartner}(1996)}]{1996ApJ...470..551D}
{Draine}, B.~T. \& {Weingartner}, J.~C. 1996, \apj, 470, 551

\bibitem[{{Fitzpatrick} \& {Massa}(1990)}]{1990ApJS...72..163F}
{Fitzpatrick}, E.~L. \& {Massa}, D. 1990, \apjs, 72, 163

\bibitem[{{Fitzpatrick} \& {Massa}(2005)}]{2005AJ....130.1127F}
{Fitzpatrick}, E.~L. \& {Massa}, D. 2005, \aj, 130, 1127

\bibitem[{{Foreman-Mackey} {et~al.}(2013){Foreman-Mackey}, {Hogg}, {Lang}, \&
  {Goodman}}]{emcee}
{Foreman-Mackey}, D., {Hogg}, D.~W., {Lang}, D., \& {Goodman}, J. 2013, PASP,
  125, 306

\bibitem[{{Gold}(1952)}]{Gold1952}
{Gold}, T. 1952, \mnras, 112, 215

\bibitem[{{Gordon} {et~al.}(2009){Gordon}, {Cartledge}, \&
  {Clayton}}]{2009ApJ...705.1320G}
{Gordon}, K.~D., {Cartledge}, S., \& {Clayton}, G.~C. 2009, \apj, 705, 1320

\bibitem[{{Hall}(1949)}]{1949Sci...109..166H}
{Hall}, J.~S. 1949, Science, 109, 166

\bibitem[{{Hensley} \& {Draine}(2023)}]{2023ApJ...948...55H}
{Hensley}, B.~S. \& {Draine}, B.~T. 2023, \apj, 948, 55

\bibitem[{{Herranen} {et~al.}(2021){Herranen}, {Lazarian}, \&
  {Hoang}}]{2021ApJ...913...63H}
{Herranen}, J., {Lazarian}, A., \& {Hoang}, T. 2021, ApJ, 913, 63

\bibitem[{{Hiltner}(1949)}]{1949Sci...109..165H}
{Hiltner}, W.~A. 1949, Science, 109, 165

\bibitem[{Hoang(2017)}]{Hoang.2017SNe}
Hoang, T. 2017, ApJ, 836, 13

\bibitem[{{Hoang}(2025)}]{2025ApJ...994..115H}
{Hoang}, T. 2025, \apj, 994, 115

\bibitem[{Hoang \& Lazarian(2008)}]{HoangLaz.2008}
Hoang, T. \& Lazarian, A. 2008, \mnras, 388, 117

\bibitem[{{Hoang} \& {Lazarian}(2016{\natexlab{a}})}]{HoangLaz.2016a}
{Hoang}, T. \& {Lazarian}, A. 2016{\natexlab{a}}, ApJ, 831, 159

\bibitem[{{Hoang} \& {Lazarian}(2016{\natexlab{b}})}]{HoangLaz.2016b}
{Hoang}, T. \& {Lazarian}, A. 2016{\natexlab{b}}, ApJ, 821, 91

\bibitem[{{Hoang} {et~al.}(2015){Hoang}, {Lazarian}, \&
  {Andersson}}]{hoang2015}
{Hoang}, T., {Lazarian}, A., \& {Andersson}, B.-G. 2015, \mnras, 448, 1178

\bibitem[{Hoang {et~al.}(2014)Hoang, Lazarian, \& Martin}]{HoangLazMartin.2014}
Hoang, T., Lazarian, A., \& Martin, P.~G. 2014, ApJ, 790, 6

\bibitem[{{Hoang} {et~al.}(2014){Hoang}, {Lazarian}, \&
  {Martin}}]{2014ApJ...790....6H}
{Hoang}, T., {Lazarian}, A., \& {Martin}, P.~G. 2014, ApJ, 790, 6

\bibitem[{{Hoang} {et~al.}(2021){Hoang}, {Tram}, {Lee}, {Diep}, \&
  {Ngoc}}]{2021ApJ...908..218H}
{Hoang}, T., {Tram}, L.~N., {Lee}, H., {Diep}, P.~N., \& {Ngoc}, N.~B. 2021,
  ApJ, 908, 218

\bibitem[{{Hoang} {et~al.}(2022){Hoang}, {Tram}, {Minh Phan}, {Giang},
  {Phuong}, \& {Dieu}}]{2022AJ....164..248H}
{Hoang}, T., {Tram}, L.~N., {Minh Phan}, V.~H., {et~al.} 2022, AJ, 164, 248

\bibitem[{{Jones} \& {Spitzer}(1967)}]{1967ApJ...147..943J}
{Jones}, R.~V. \& {Spitzer}, Jr., L. 1967, \apj, 147, 943

\bibitem[{{Kim} \& {Martin}(1995)}]{kim1995}
{Kim}, S.-H. \& {Martin}, P.~G. 1995, \apj, 444, 293

\bibitem[{{Lallement} {et~al.}(2014){Lallement}, {Vergely}, {Valette},
  {Puspitarini}, {Eyer}, \& {Casagrande}}]{lallement2014}
{Lallement}, R., {Vergely}, J.~L., {Valette}, B., {et~al.} 2014, \aap, 561, A91

\bibitem[{{Lazarian}(1997)}]{1997MNRAS.288..609L}
{Lazarian}, A. 1997, \mnras, 288, 609

\bibitem[{{Lazarian} {et~al.}(2015){Lazarian}, {Andersson}, \&
  {Hoang}}]{2015psps.book...81L}
{Lazarian}, A., {Andersson}, B.-G., \& {Hoang}, T. 2015, in Polarimetry of
  Stars and Planetary Systems, ed. L.~{Kolokolova}, J.~{Hough}, \& A.-C.
  {Levasseur-Regourd}, 81

\bibitem[{{Lazarian} \& {Draine}(2000)}]{2000ApJ...536L..15L}
{Lazarian}, A. \& {Draine}, B.~T. 2000, \apjl, 536, L15

\bibitem[{{Lazarian} \& {Hoang}(2007{\natexlab{a}})}]{lazarian2007}
{Lazarian}, A. \& {Hoang}, T. 2007{\natexlab{a}}, \mnras, 378, 910

\bibitem[{{Lazarian} \& {Hoang}(2007{\natexlab{b}})}]{2007MNRAS.378..910L}
{Lazarian}, A. \& {Hoang}, T. 2007{\natexlab{b}}, \mnras, 378, 910

\bibitem[{{Lazarian} \& {Hoang}(2021)}]{2021ApJ...908...12L}
{Lazarian}, A. \& {Hoang}, T. 2021, ApJ, 908, 12

\bibitem[{{Lazarian} \& {Roberge}(1997)}]{1997ApJ...484..230L}
{Lazarian}, A. \& {Roberge}, W.~G. 1997, \apj, 484, 230

\bibitem[{{Lee} {et~al.}(2020){Lee}, {Hoang}, {Le}, \&
  {Cho}}]{2020ApJ...896...44L}
{Lee}, H., {Hoang}, T., {Le}, N., \& {Cho}, J. 2020, ApJ, 896, 44

\bibitem[{{Maconi} {et~al.}(2023){Maconi}, {Soler}, {Reissl}, {Girichidis},
  {Klessen}, {Hennebelle}, {Molinari}, {Testi}, {Smith}, {Sormani}, {Teh}, \&
  {Traficante}}]{2023MNRAS.523.5995M}
{Maconi}, E., {Soler}, J.~D., {Reissl}, S., {et~al.} 2023, \mnras, 523, 5995

\bibitem[{{Martin} {et~al.}(1999){Martin}, {Clayton}, \&
  {Wolff}}]{1999ApJ...510..905M}
{Martin}, P.~G., {Clayton}, G.~C., \& {Wolff}, M.~J. 1999, \apj, 510, 905

\bibitem[{{Mathis} {et~al.}(1983){Mathis}, {Mezger}, \&
  {Panagia}}]{1983A&A...128..212M}
{Mathis}, J.~S., {Mezger}, P.~G., \& {Panagia}, N. 1983, A\&A, 500, 259

\bibitem[{{Mathis} {et~al.}(1977){Mathis}, {Rumpl}, \&
  {Nordsieck}}]{1977ApJ...217..425M}
{Mathis}, J.~S., {Rumpl}, W., \& {Nordsieck}, K.~H. 1977, ApJ, 217, 425

\bibitem[{{Mauron} \& {Huggins}(2010)}]{mauron2010}
{Mauron}, N. \& {Huggins}, P.~J. 2010, \aap, 513, A31

\bibitem[{{Papoular}(2018)}]{2018MNRAS.479.1685P}
{Papoular}, R. 2018, \mnras, 479, 1685

\bibitem[{{Purcell}(1979)}]{1979ApJ...231..404P}
{Purcell}, E.~M. 1979, ApJ, 231, 404

\bibitem[{{Ronchi} {et~al.}(2021){Ronchi}, {Graber}, {Garcia-Garcia}, {Rea}, \&
  {Pons}}]{ronchi2021}
{Ronchi}, M., {Graber}, V., {Garcia-Garcia}, A., {Rea}, N., \& {Pons}, J.~A.
  2021, \apj, 916, 100

\bibitem[{{Serkowski}(1971)}]{serkowski1971}
{Serkowski}, K. 1971, in IAU Colloquium 15: New Directions and New Frontiers in
  Variable Star Research, ed. W.~{Strohmeier}, 11

\bibitem[{{Serkowski}(1973)}]{1973IAUS...52..145S}
{Serkowski}, K. 1973, in Interstellar Dust and Related Topics, ed. J.~M.
  {Greenberg} \& H.~C. {van de Hulst}, Vol.~52, 145

\bibitem[{{Siebenmorgen} {et~al.}(2014){Siebenmorgen}, {Voshchinnikov}, \&
  {Bagnulo}}]{2014A&A...561A..82S}
{Siebenmorgen}, R., {Voshchinnikov}, N.~V., \& {Bagnulo}, S. 2014, \aap, 561,
  A82

\bibitem[{{Sobey} {et~al.}(2019){Sobey}, {Bilous}, {Grie{\ss}meier}, {Hessels},
  {Karastergiou}, {Keane}, {Kondratiev}, {Kramer}, {Michilli}, {Noutsos},
  {Pilia}, {Polzin}, {Stappers}, {Tan}, {van Leeuwen}, {Verbiest},
  {Weltevrede}, {Heald}, {Alves}, {Carretti}, {En{\ss}lin}, {Haverkorn},
  {Iacobelli}, {Reich}, \& {Van Eck}}]{sobey2019}
{Sobey}, C., {Bilous}, A.~V., {Grie{\ss}meier}, J.~M., {et~al.} 2019, \mnras,
  484, 3646

\bibitem[{{Stil} {et~al.}(2009){Stil}, {Wityk}, {Ouyed}, \&
  {Taylor}}]{stil2009}
{Stil}, J., {Wityk}, N., {Ouyed}, R., \& {Taylor}, A.~R. 2009, \apj, 701, 330

\bibitem[{{Tram} \& {Hoang}(2022)}]{2022FrASS...9.3927T}
{Tram}, L.~N. \& {Hoang}, T. 2022, Front. astron. space sci., 9, 923927

\bibitem[{{Tram} {et~al.}(2025){Tram}, {Hoang}, {Lazarian}, {Seifried},
  {Andersson}, {Pillai}, {Truong}, {Diep}, \&
  {Fanciullo}}]{2025A&A...703A.192T}
{Tram}, L.~N., {Hoang}, T., {Lazarian}, A., {et~al.} 2025, \aap, 703, A192

\bibitem[{{Weaver} {et~al.}(1977){Weaver}, {McCray}, {Castor}, {Shapiro}, \&
  {Moore}}]{1977ApJ...218..377W}
{Weaver}, R., {McCray}, R., {Castor}, J., {Shapiro}, P., \& {Moore}, R. 1977,
  \apj, 218, 377

\bibitem[{{Whittet}(2003)}]{whittet2003}
{Whittet}, D. C.~B. 2003, Dust in the galactic environment - 2:nd ed. (Dust in
  the galactic environment Institute of Physics Publishing, 390 p.)

\bibitem[{{Whittet} {et~al.}(1992){Whittet}, {Martin}, {Hough}, {Rouse},
  {Bailey}, \& {Axon}}]{1992ApJ...386..562W}
{Whittet}, D.~C.~B., {Martin}, P.~G., {Hough}, J.~H., {et~al.} 1992, \apj, 386,
  562

\bibitem[{{Wilking} {et~al.}(1980){Wilking}, {Lebofsky}, {Kemp}, {Martin}, \&
  {Rieke}}]{wilking1980}
{Wilking}, B.~A., {Lebofsky}, M.~J., {Kemp}, J.~C., {Martin}, P.~G., \&
  {Rieke}, G.~H. 1980, \apj, 235, 905

\bibitem[{{Wilking} {et~al.}(1982){Wilking}, {Lebofsky}, \&
  {Rieke}}]{wilking1982}
{Wilking}, B.~A., {Lebofsky}, M.~J., \& {Rieke}, G.~H. 1982, \aj, 87, 695

\bibitem[{{Zuckerman} {et~al.}(2012){Zuckerman}, {Melis}, {Rhee}, {Schneider},
  \& {Song}}]{2012ApJ...752...58Z}
{Zuckerman}, B., {Melis}, C., {Rhee}, J.~H., {Schneider}, A., \& {Song}, I.
  2012, \apj, 752, 58

\bibitem[{{Zuo} {et~al.}(2021){Zuo}, {Li}, \& {Zhao}}]{2021ApJS..257...63Z}
{Zuo}, W., {Li}, A., \& {Zhao}, G. 2021, \apjs, 257, 63

\end{thebibliography}

\appendix
\section{Geometrical factor and magnetic susceptibility of grain} \label{app:grain_phys}
For a grain with an eccentricity of $e_{\rm m} = \sqrt{1-(a/b)^{2}}$, the geometrical factor as \cite{1997MNRAS.288..609L}
\begin{equation}
    \Gamma_{\parallel}(e_{\rm m}) = \frac{3}{16}\left\{3 + 4(1-e_{\rm m}^{2})g_{\rm m}(e_{\rm m}) - e_{\rm m}^{-2}[1 - (1-e_{\rm m}^{2})^{2}g_{\rm m}(e_{\rm m})]\right\}
\end{equation}
with $g_{\rm m}(e_{\rm m})=\frac{1}{2e_{\rm m}}\ln\left(\frac{1+e_{\rm m}}{1-e_{\rm m}}\right)$. For an oblate grain of $b/a=1.4$ in this work, $s=a/b=0.7$, $e_{\rm m}=0.7$ and $\Gamma_{\parallel} \simeq 0.78$.

The gas damping timescale is (see e.g. \citealt{2022AJ....164..248H})
\begin{equation}
    \tau_{\rm gas} = \frac{3}{4\sqrt{\pi}}\frac{I_{\parallel}}{1.2n_{\rm H}m_{\rm H}v_{\rm th}a^{4}\Gamma_{\parallel}}
\end{equation}
where $I_{\parallel} = \frac{8}{15}\rho a^{5}$ is the inertial momentum of grain and $v_{\rm th} = \sqrt{2k_{\rm B}T_{\rm gas}/m_{\rm H}}$ is the thermal velocity of gas.

The timescale for the DG relaxation is (see, e.g. \citealt{HoangLaz.2016a})
\begin{equation}
    \tau_{\rm DG} = \frac{I_{\parallel}}{K(\rm \omega)V B^{2}}
\end{equation}
where $V=4/3\pi a^{3}$, $B$ is the magnetic field strength and $K(\omega)$ is defined as

\begin{equation}
    \begin{split}
    K(\omega) &=\frac{\chi(0)\tau_{\rm sp}}{\left[1+\left(\omega \tau_{\rm el}/2\right)^{2}\right]^{2}} \\
    &\simeq 5.2\times 10^{-11}N_{\rm cl}\phi_{sp}\left(\frac{p}{5.5}\right)^{2}\left(\frac{T_{\rm dust}}{10\,\rm K}\right)^{-1}e^{\frac{0.011{\,\rm K} N_{\rm cl}}{T_{\rm dust}}}\left[1+\left(\frac{\omega\tau_{sp}}{2}\right)^{2}\right]^{-2}
    \end{split}
\end{equation}
with $\chi(0)=0.026N_{\rm cl}\phi_{\rm sp}\left(\frac{20\, \rm K}{T_{\rm dust}}\right)$ the magnetic susceptibility of grain at $\omega=0$, and $\tau_{\rm sp}=10^{-9} e^{0.011{\rm K}N_{\rm cl}/T_{\rm dust}}\,$s the re-magnetisation timescale by thermal fluctuation. $N_{\rm cl}=1$ and $\phi_{\rm sp}=1/7$ represent the paramagnetic grain. These values change (e.g. higher $N_{\rm cl}$) for super-paramagnetic grains.

\section{Polarisation-average extinction and polarisation cross-sections} \label{sec:cross-sections}
Following \cite{2009ApJ...696....1D}, when the incident electric field $\boldsymbol{E}$ inclines with an angle $\xi$ with the symmetric axis $\boldsymbol{a}_{1}$ of an irregular grain, the extinction cross-section is
\begin{equation}
    C_{\rm ext} = C_{\rm ext,\mathbf{E\parallel a}}\cos^{2}\xi + C_{\rm ext, \mathbf{E\perp a}}\sin^{2}\xi 
\end{equation}
The total cross-section of an aligned grain with alignment efficiency $f(a)$ can be estimated by averaging the angle $\xi$. The average of this angle can be decomposed as $\langle \cos^{2}\xi \rangle$ = $\langle \cos^{2}\xi_{\parallel} \rangle$ + $\langle \cos^{2}\xi_{\perp} \rangle$ and $\langle \sin^{2}\xi \rangle$ = $\langle \sin^{2}\xi_{\parallel} \rangle$ + $\langle \sin^{2}\xi_{\perp} \rangle$ with 
\begin{equation}
    \begin{split}
        &\langle \cos^{2}\xi_{\parallel} \rangle = \frac{1-f}{3} + f\sin^{2}\psi ~~~ {\rm and} ~~~ \langle \sin^{2}\xi_{\parallel}\rangle = 1- \langle \cos^{2}\xi_{\parallel} \rangle, \\
        &\langle \cos^{2}\xi_{\perp} \rangle = \frac{1-f}{3} ~~~~~~~~~~~~~~~~~~~ {\rm and} ~~~ \langle \sin^{2}\xi_{\perp}\rangle = 1- \langle \cos^{2}\xi_{\perp} \rangle.
    \end{split}
\end{equation}
The first term is for randomised grains, while the second term is for aligned grains with the angle $\psi$ between the magnetic field and the line of sight. The total polarisation-average extinction cross-section is
\begin{equation}
    \begin{split}
        \langle C_{\rm ext} \rangle &= \frac{1}{3}(2C_{\rm ext,\mathbf{E\perp a}} + C_{\rm ext, \mathbf{E\parallel a}}) + \frac{1}{2}f(a)\left(\cos^{2}\psi - \frac{1}{3}\right)(C_{\rm ext,\mathbf{E\perp a}} - C_{\rm ext, \mathbf{E\parallel a}}) \\
        &= C^{\rm ran}_{\rm ext} + f(a)C_{\rm pol}\left(\frac{2}{3}-\sin^{2}\psi\right)
        \end{split}
\end{equation}
with $C^{\rm ran}_{\rm ext}=1/3(2C_{\rm ext, \boldsymbol{E}\perp \boldsymbol{a}} + C_{\rm ext,\boldsymbol{E}\parallel \boldsymbol{a}})$ the extinction cross-section of randomly aligned grains, and $C_{\rm pol}=0.5(C_{\rm ext,\mathbf{E\parallel a}}\cos^{2}\xi - C_{\rm ext, \mathbf{E\perp a}}\sin^{2}\xi)  = 0.5f(a)\sin^{2}\psi(C_{\rm ext,\mathbf{E\perp a}} - C_{\rm ext, \mathbf{E\parallel a}})$ the polarisation cross-section.

\section{Spectrum of starlight polarisation}
\begin{figure*}
    \centering
    \includegraphics[width=1.0\linewidth]{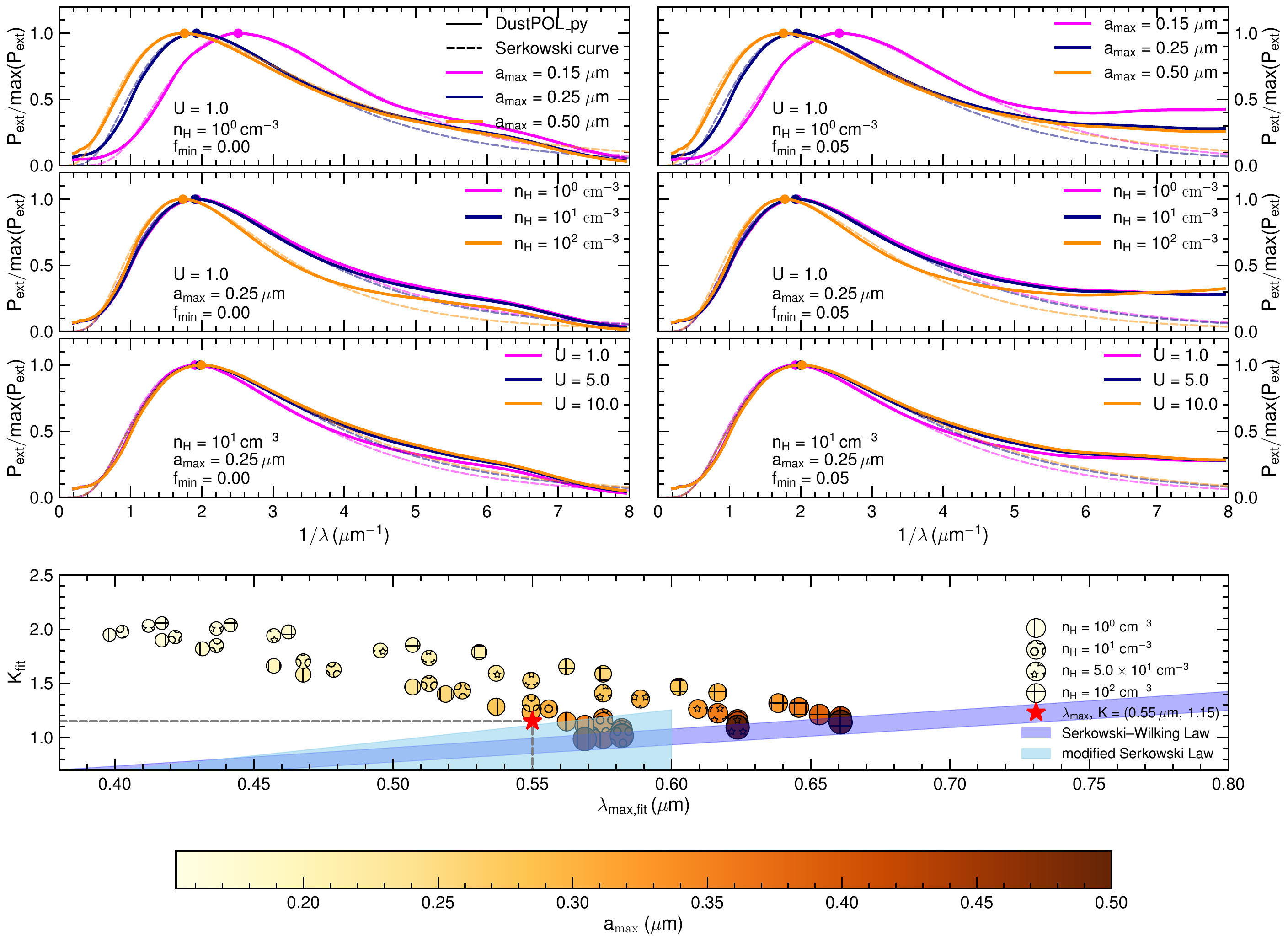}
    \caption{Top panels: Spectrum calculated from our model (solid lines) with the fitting from the Serkowski curves (dashed lines). The top left panels are for $f_{\rm min}=0$ while the top right panels are for $f_{\rm min}>0$. Bottom panel: $K-\lambda_{\rm max}$ relation for different values of $a_{\rm max}$ and $n_{\rm H}$ with $U=1$. The variation in the $a_{\rm max}$ values is colour-coded and proportional to the symbol size. In this plot, the alignment degree by DG alignment $f_{\rm min}$ are set manually to study the effect of the alignment of small grains.}
    \label{fig:physical_variations}
\end{figure*}

To show the effect of DG alignment, we manually assign the values of $f_{\rm min}$ in Figure \ref{fig:physical_variations}. The top panels illustrate the changes in starlight polarisation spectra from the near-IR to the far-UV range, associated with varying radiation field values, gas densities, and maximum grain sizes for $f_{\rm min}=0$ (left panels) and $f_{\rm min}=0.05$ (right panels). Each spectrum is then fitted by a Serkowski curve, depicted by the dashed line. The peak wavelength ($\lambda_{\rm max}$) is primarily affected by the maximum grain size and gas density, whereas the spectral width is governed by all the radiation field, the gas density and the maximum grain size. Compared to the Serkowski curve, a lower gas density combined with higher radiation can cause an increase at higher wavenumbers. However, this excess disappears at the maximum wavenumbers when $f_{\rm min}=0$, but remains significant when $f_{\rm min}>0$.

The bottom panel of Figure \ref{fig:physical_variations} illustrates the $K-\lambda_{\rm max}$ relations induced by the Serkowski fitting shown in the top panel for $f_{\rm min}=0$ and $U=1$. For the same $a_{\rm max}$, $K$ is positively correlated with $\lambda_{\rm max}$ with increasing $n_{\rm H}$. The slope of this relation in the mid-visible to IR ($\lambda_{\rm max}>0.55\,\mu$m) is seen to follow nicely the Serkowki-Wilking law (Equation \ref{eq:SW_law} and blue line). While from the mid-visible to UV ($\lambda_{\rm max}<0.55\,\mu$m) the slope is slightly steeper and closer to the modified Serkowski law (Equation \ref{eq:modS_law} and the cyan line). 
For a given $n_{\rm H}$, $K$ is negatively correlated with $\lambda_{\rm max}$ with increasing $a_{\rm max}$. This is because there are more grains to be aligned by RATs ($a>a_{\rm align}$, i.e. broader the size distribution of aligned grains).

\section{Magnetic field in Per OB3 super-bubble} \label{sec:Bfield_perOB3}
\begin{figure}
    \centering
    \includegraphics[width=1.0\linewidth]{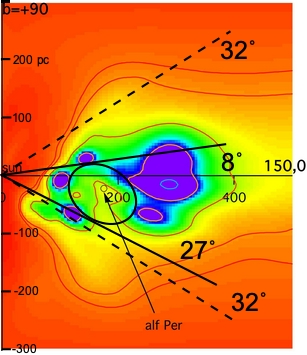}
    \caption{Observed extent of the Per OB3 ($\alpha$ Per) bubble in the direction of \textit{lon}=150$^\circ$.  Fitting an oval to the inside of the high extinction regions around Per OB3 from \citet{lallement2014} we find an opening angle of $\sim$35$^\circ$, while the CO (J=1-0) data \citep{bhatt2000} yields an opening angle of $\sim$65$^\circ$ (adapted with permission from \citealt{lallement2014}).}
    \label{fig:PerOB3_ext}
\end{figure}

To evaluate the relative contribution of EUV driven RAT alignment and Davis-Greenstein alignment of very small grains, it is useful to constrain the magnetic field in the materials, as the latter needs a strong field to be efficient. 
\citet{lallement2014} used extinction data to map the local ISM, including the region around Per OB3, showing a cavity of some 35$^\circ$ extent on the sky (Figure \ref{fig:PerOB3_ext}). \citet{bhatt2000}, in turn, used CO (J=1-0) observations and found a somewhat larger extent of the cavity around the association of $\sim$50-65$^\circ$.

\begin{table*}[]
    \centering
    \caption{Faraday rotation sources}
    \begin{tabular}{l|c c c c c}
         \hline
         \hline
         Target & RA & Dec & Offset & d$^{a}$ & B$^{b}$  \\
         {}     & {J2000} & {J2000}  & [$^\circ$]  & [kpc] & [G] \\
         \hline
    B0450+55 & 04:54:07.71 & +55:43:41.5 & 14.2 & 0.64 & 0.49$\pm$0.01\\
    J0636+5129 & 06:36:04.85 & +51:29:00.0 & 29.3 & 0.21 & -0.24$\pm$0.01\\
    J0645+5158 & 06:45:59.08 & +51:58:14.9 & 30.7 & 0.67 & -0.13$\pm$0.01\\
    B0301+19 & 03:04:33.12 & +19:32:51.4 & 30.7 & 0.74 & -0.66$\pm$0.01\\
    B0809+74 & 08:14:59.50 & +74:29:05.7 & 37.7 & 0.37 & -2.98$\pm$0.02\\
    B1322+83 & 13:21:46.18 & +83:23:38.9 & 45.8 & 0.98 & -2.19$\pm$0.01\\
    J0621+1002 & 06:21:22.11 & +10:02:38.7 & 53.5 & 0.14 & 1.79$\pm$0.01\\
    B0656+14 & 06:59:48.18 & +14:14:21.5 & 55.7 & 0.16 & 1.99$\pm$0.01\\
    J1012+5307 & 10:12:33.43 & +53:07:02.6 & 57.5 & 0.81 & 0.41$\pm$0.01\\
    J0030+0451 & 00:30:27.43 & +04:51:39.7 & 58.4 & 0.35 & 0.33$\pm$0.02\\
    B0823+26 & 08:26:51.43 & +26:37:22.8 & 60.4 & 0.31 & 0.32$\pm$0.01\\
    B1112+50 & 11:15:38.40 & +50:30:12.3 & 66.1 & 0.92 & 0.32$\pm$0.01\\
         \hline
    \end{tabular}

   \tablefoot{a: From \citet{ronchi2021}, b: From \citet{sobey2019}}
   \label{tab:FR_info}
\end{table*}

\begin{figure}
    \centering
    \includegraphics[width=1.0\linewidth]{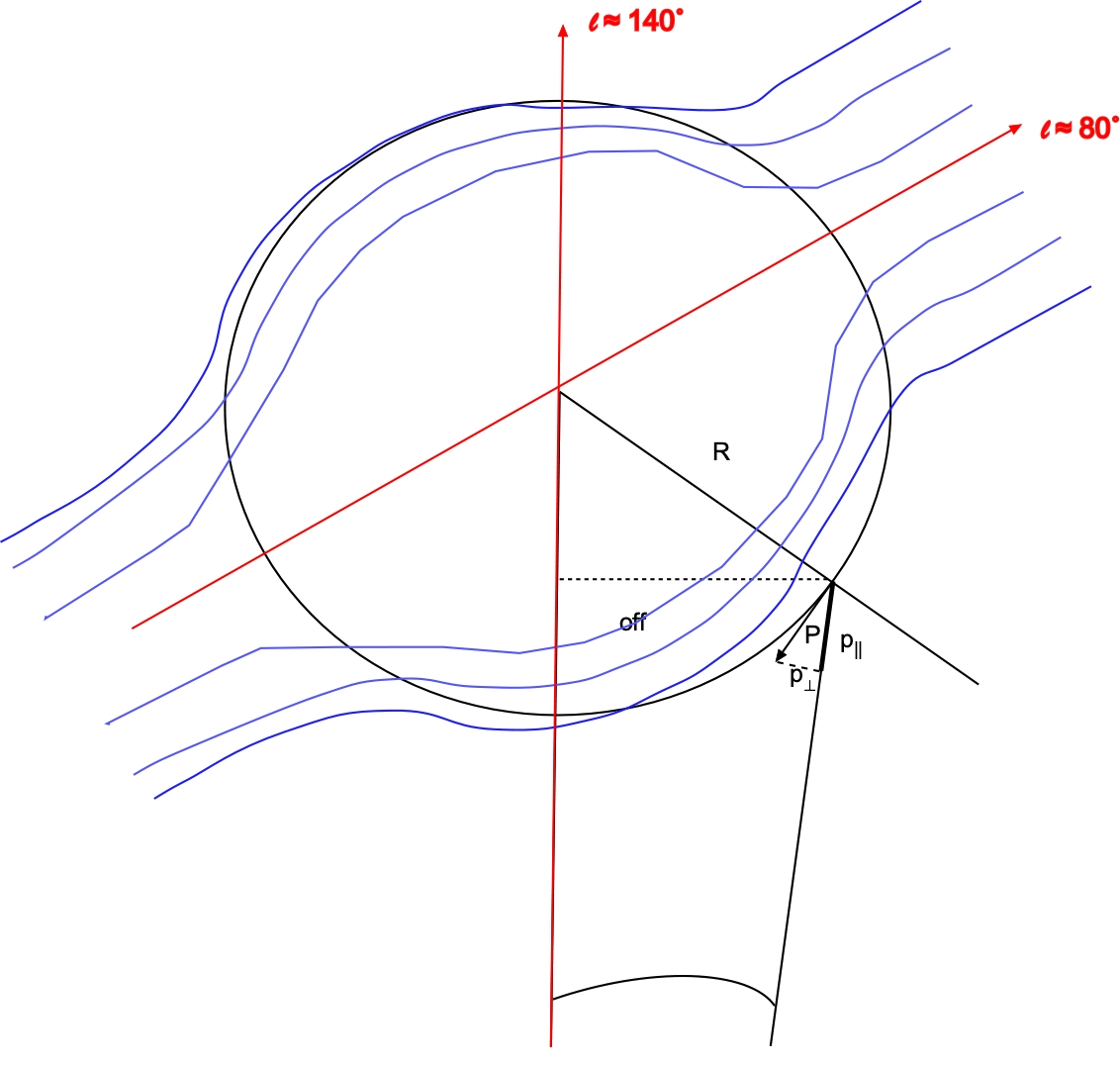}
    \caption{Toy model of the super-bubble.}
    \label{fig:superbubble_toymodel}
\end{figure}

We used the magnetic field measurements using Faraday rotation (and dispersion measures) from \citet{sobey2019} combined with the pulsar distances of \citet{ronchi2021} to estimate the magnetic field strength in the Per OB3 super-bubble.  
To avoid background contamination, we restricted the sample to pulsars with d$<$1 kpc (and within 70$^\circ$ of the centre of the association), which yields a sample of 12 pulsars with measurements of the interstellar magnetic field.  Using a very simple model of an entrenched magnetic field in the wall of the bubble (Figure \ref{fig:superbubble_toymodel} we can fit the field strength. 
If we assume that the B-field is blown into the shell of the bubble \citep[e.g.][]{stil2009} and we assume that it originates as the average local [large-scale] Galactic field (directed at \textit{lon}$\sim$80$^\circ$ \citealt{crutcher2003}).  Figure \ref{fig:superbubble_toymodel} illustrates our toy model, in which the field lines are mostly in the plane of the sky.

\begin{figure}
    \centering
    \includegraphics[width=1.0\linewidth]{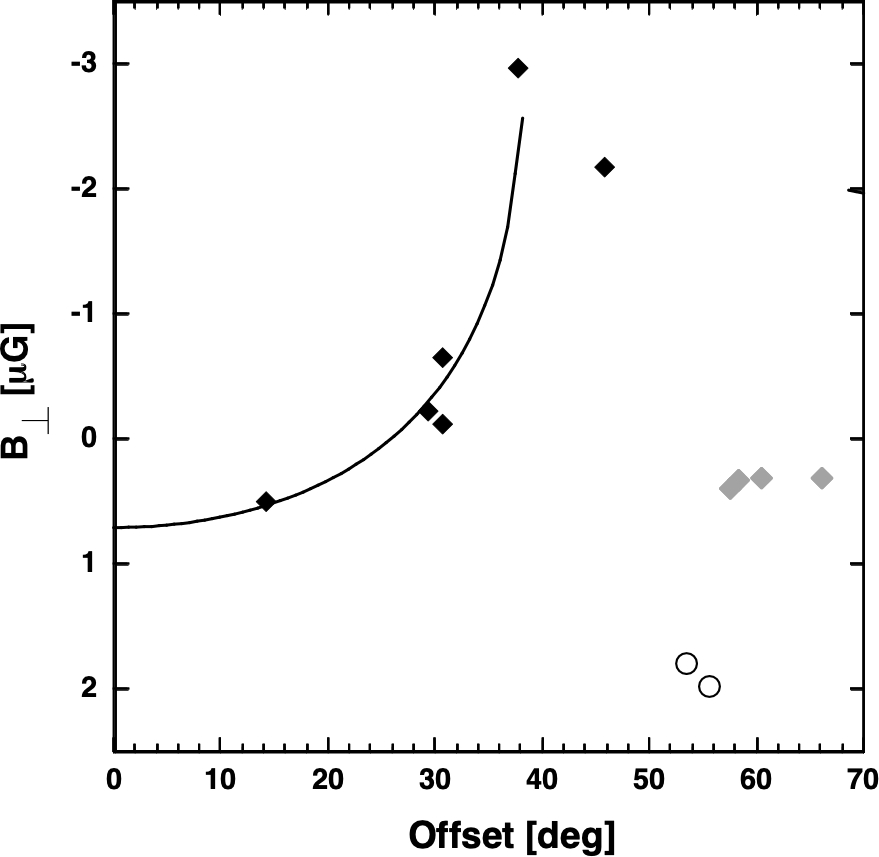}
    \caption{Fitted magnetic field strengths for the Pe OB3 bubble.  While Faraday rotation polarisation traces the magnetic field parallel to the line of sight, we derive its component perpendicular to the LOS because of the assumed source geometry.  The uncertainties for each data point from \citet{sobey2019} are smaller than the plotting symbols.}
    \label{fig:superbubble_fit}
\end{figure}

Because the model assumes a spherical structure, we - based on the results from \citet{lallement2014} and \citet{bhatt2000} restrict the Faraday rotation sample in our fit further to include only those sources with 50$^\circ$ of the centre of the bubble (leaving 6 sources in the fitting). Figure \ref{fig:superbubble_fit} shows the best-fit model, where the targets shown with open symbols are located at d$<$200 pc (approximate distance of the Per OB3 association), and the targets shown with gray diamonds are located at larger distances (but less than 1 kpc).  Although simple and based on a limited number of data points, the fit yields a reasonable field strength of B$_{||}\approx$ 3.3$\mu$G (assuming a fore- or background component accounting for the non-zero value at the centre of the bubble), and a bubble extent of $\sim$38.

 If we assume that the polarisation is in the POS for the whole bubble then we get a really good fit to the data. This is with the background pulsars with $200 < d < 1000\,\rm pc$. Taking into account the projection effect, the Bfield strength is about 5$\,\mu$m. 

\section{Super-Serkowski spectrum within super-bubble condition for a higher gas density} 
\begin{figure}
    \centering
    \includegraphics[width=1.0\linewidth]{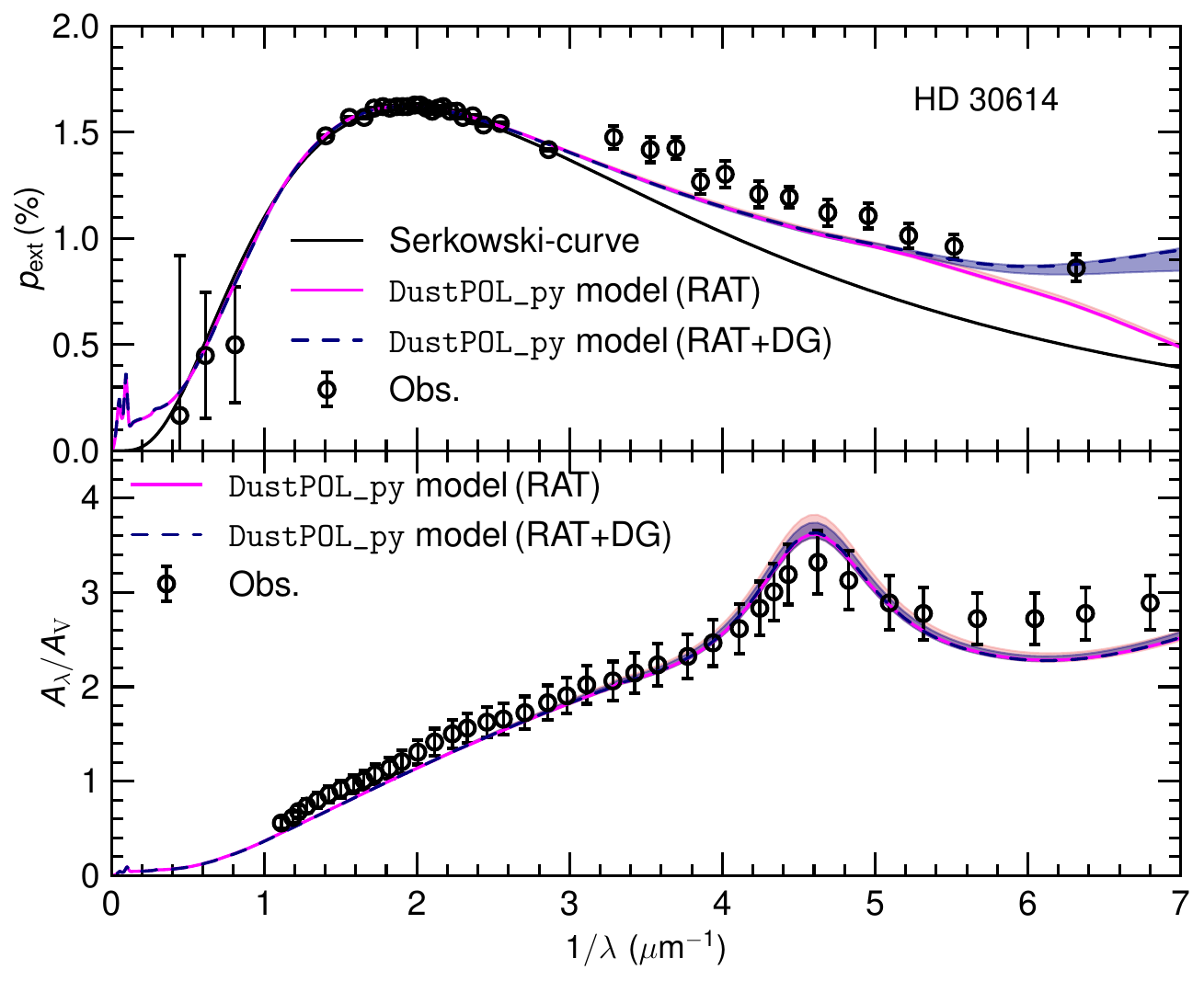}
    \caption{Similar to Figure \ref{fig:HD30614_EUV} but with $n_{\rm H}=30\,\rm cm^{-3}$ for HD 30614. Compared to the case of lower gas density, the degree of polarisation towards the highest wavenumber becomes lower for denser gas because of the gas randomisation effect.}
    \label{fig:supSep_bubble_unreal}
\end{figure}

Figure \ref{fig:supSep_bubble_unreal} shows the best fit for HD 30614 with $n_{\rm H}=30\,\rm cm^{-3}$. Compared to Figure \ref{fig:HD30614_EUV}, the degree of mid-UV polarisation tends to be lower because collisional damping for the DG alignment is more important in denser gas. Therefore, spectropolarimetric observations at mid-UV and far-UV are required.

\end{document}